\documentclass[12pt]{article}
\usepackage[T1]{fontenc}
\usepackage{geometry}
\usepackage[utf8]{inputenc}
\usepackage{float}
\usepackage{amsmath}
\usepackage{amssymb}
\usepackage{graphicx}

\makeatletter

\floatstyle{ruled}
\newfloat{algorithm}{tbp}{loa}
\providecommand{\algorithmname}{Algorithm}
\floatname{algorithm}{\protect\algorithmname}

\usepackage{stackrel}

\@ifundefined{showcaptionsetup}{}{%
 \PassOptionsToPackage{caption=false}{subfig}}
\usepackage{subfig}
\makeatother

\begin{document}

\title{Shape-Constrained Density Estimation Via Optimal Transport}

\author{Ryan Cumings-Menon\thanks{Department of Economics, University of Illinois at Urbana-Champaign,
\textit{Address:} 214 David Kinley Hall, 1407 West Gregory Drive,
Urbana, IL 61801, USA. \textit{E-mail address:} cumings2@illinois.edu.}}
\maketitle
\begin{abstract}
Constraining the maximum likelihood density estimator to satisfy a sufficiently
strong constraint, $\log-$concavity being a common example, has the
effect of restoring consistency without requiring additional parameters.
Since many results in economics require densities to satisfy a regularity 
condition, these estimators are also attractive for the structural
estimation of economic models. In all of the examples of regularity conditions provided by Bagnoli
and Bergstrom (2005) and Ewerhart (2013), $\log-$concavity is sufficient
to ensure that the density satisfies the required conditions. However,
in many cases $\log-$concavity is far from necessary, and it has
the unfortunate side effect of ruling out sub-exponential tail behavior. 

In this paper, we use optimal transport to formulate a shape constrained
density estimator. We initially describe the estimator using a $\rho-$concavity
constraint. In this setting we provide results on consistency, asymptotic
distribution, convexity of the optimization problem defining the estimator,
and formulate a test for the null hypothesis that the population density
satisfies a shape constraint. Afterward, we provide sufficient conditions
for these results to hold using an arbitrary shape constraint. This
generalization is used to explore whether the California Department
of Transportation's decision to award construction contracts with
the use of a first price auction is cost minimizing. We estimate the
marginal costs of construction firms subject to Myerson's (1981) regularity
condition, which is a requirement for the first price reverse auction
to be cost minimizing. The proposed test fails to reject that the
regularity condition is satisfied. 


\end{abstract}

\section{Introduction}

Nonparametric density estimation has the advantage over its parametric
counterparts of not requiring the underlying population density to
belong to a specific family. In the case of distribution functions,
Kiefer and Wolfowitz (1956) showed that the empirical distribution
function is a maximum likelihood estimator; however, attempting to
use this distribution function to directly define a nonparametric
density estimate results in a series of point masses located at each
of the datapoints. Grenander (1956) provided the first example of
a shape constrained density estimator as a way to extricate the maximum
likelihood estimator from this ``Dirac catastrophe.'' Specifically,
he showed that maximizing the likelihood function subject to a monotonicity
constraint on the estimator results in density estimates without point
masses. 

A great deal of progress was made in subsequent decades by adding
penalty terms to the maximum likelihood objective function to restore
the parsimony of the density estimator; for example, see (Silverman,
1986). Parzen (1962) also showed that kernel density estimators resulted
in consistent density estimators and derived the rates of convergence.
Unlike Grenander's (1956) approach, the performance of maximum penalized
likelihood estimators and kernel density estimators is highly dependent
on the specification of penalty terms and bandwidths, respectively,
which can be difficult to choose. 

Partly for this reason, recently there has been a renewed interest
in ensuring parsimony of the maximum likelihood density estimator
through conditioning on the information provided by the shape of the
underlying density. In particular, significant progress has been made
on the maximum likelihood density estimator subject to the constraint
that the logarithm of the density is a concave function, which defines
a $\log-$concave density (Dümbgen and Rufibach, 2009; Cule, Samworth,
and Stewart, 2010; Kim and Samworth, 2016). 

One early pioneer on the advantages of $\log-$concavity for both
statistical testing as well as estimation was Karlin (1968). Suppose
the distribution function $F:\mathbb{R}\rightarrow[0,1]$ has a density
function denoted by $f:\mathbb{R}\rightarrow\mathbb{R}_{+}.$ Some
examples of $\log-$concavity's many implications include that the
density $f(x-\theta)$ has a monotonic likelihood ratio if and only
if $f(\cdot)$ is $\log-$concave, products and convolutions between
$\log-$concave densities are $\log-$concave, and that the hazard
function of the $\log-$concave density $f(x),$ defined by $f(x)/(1-F(x)),$
is increasing. Bagnoli and Bergstrom (2005) also provide a survey
of economic models in which $\log-$concavity of a density is a sufficient
condition for the existence or uniqueness of an equilibrium. Chen
and Samworth (2013) as well as Dümbgen, Samworth, and Schuhmacher
(2011) provide tests for a population density satisfying $\log-$concavity,
and Carroll, Delaigle, and Hall (2011) provide a test for a population
density satisfying a more general set of shape constraints.

A wide variety of the random variables in the economics literature,
such as annual income or changes in stock prices, are thought to exhibit
sub-exponential tail behavior, so a $\log-$concavity constraint would
not result in a consistent estimator in these cases. Koenker and Mizera
(2010) generalized the $\log-$concave maximum likelihood estimator
by maximizing Rényi entropy of order $\rho\in\mathbb{R}$ subject
to the $\rho-$concavity constraint, 
\begin{center}
$f\left(\alpha x_{0}+(1-\alpha)x_{1}\right)\geq\left(\alpha f(x_{0})^{\rho}+(1-\alpha)f(x_{1})^{\rho}\right)^{1/\rho},$
\par\end{center}

\noindent for all $\alpha\in[0,1].$ This estimator converges to the
maximum likelihood estimator subject to a $\log-$concavity constraint
in the limit as $\rho\rightarrow0.$\footnote{Maximum likelihood is equivalent to maximizing Shannon entropy, and
Rényi entropy of order $\rho$ converges to Shannon entropy as $\rho\rightarrow0$
.} Decreasing $\rho$ corresponds to a relaxation of this shape constraint,
so if $f(x)$ satisfies the constraint for some $\rho,$ then it also
satisfies the constraint for all $\rho'<\rho.$ Also, this constraint
is equivalent to concavity when $\rho$ is equal to one, and the cases
of $\log-$concavity and quasi-concavity can be derived in the limit
as $\rho\rightarrow0$ and $\rho\rightarrow-\infty$ respectively.
Koenker and Mizera (2010) place particular emphasis on the case in
which $\rho=-1/2,$ partly because most standard densities are $-1/2-$concave.
For example, all Student $t_{v}$ densities with $v\geq1$ satisfy
this constraint. 

$\rho-$concavity constraints provide a considerable relaxation over
$\log-$concavity constraints, while restricting the set of feasible
densities sufficiently to ensure parsimony of the density estimator.
These constraints are also sufficient conditions for many results
in economics, including the uniqueness or existence of equilibria
in a variety of models; see for examples, (Ewerhart, 2013; Bagnoli
and Bergstrom, 2005). However, in many cases the necessary and sufficient
conditions for these results are considerably weaker, so inference
and estimation based on these stronger conditions can provide misleading
results. For example, inferring whether a population density satisfies
these weaker conditions based on tests for their more restrictive
counterparts is generally not possible. Things are less straightforward
in the case of estimation because shape constraints are generally
the source of the density estimator's parsimony. However, since using
a shape constraint that is not satisfied by the population density
would not result in a consistent estimator, it is prudent to err toward
the weakest constraint that theory predicts a population density would
satisfy, or when using the estimate in a structural model, the weakest
constraint that a model requires a population density to satisfy.

For a concrete example in the economics literature, given a density
of private valuations of risk neutral agents, Myerson (1981) defined
the virtual valuations function as $x-(1-F(x))/f(x),$ and showed
that the first price auction is revenue maximizing if this function
is strictly increasing. Sufficient conditions for Myerson's (1981)
regularity condition, ordered from strongest to weakest, are $\log-$concavity,
a monotonic hazard rate, and $\rho-$concavity for $\rho>-1/2$ (Ewerhart,
2013). While the first two sufficient conditions are commonly cited
in mechanism design, they both imply exponential tail behavior. Since
it seems plausible that willingness to pay is influenced by ability
to pay, allowing valuations to have sub-exponential tail behavior
may be a reasonable modeling choice, given that wealth and income
are typically modeled with sub-exponential tails. Luckily, Myerson's
(1981) regularity condition does not exclude these densities. For
example, while the $\log-$normal density has sub-exponential tails,
it also satisfies this condition when $\sigma^{2}<2,$ which holds
in the structural model provided by Laffont, Ossard, and Vuong (1995). 

This paper provides a framework for estimating and performing inference
with shape constrained densities using regularized optimal transport
(Cuturi, 2013). This objective function has the advantage of having
an unconstrained global optimum that is a consistent density estimator,
which ameliorates the requirement that the shape constraint is the
only source of parsimony. At first we motivate the method using a
$\rho-$concavity constraint, but one of the advantages of the method
is that the estimator is consistent when this constraint is replaced
by a wide variety of alternative shape constraints. We also provide
a consistent test for whether or not a population density satisfies
a shape constraint based on comparing the objective function at the
unconstrained optimum to the constrained optimum. 

After introducing density estimation with this more general class
of shape constraints, we use the proposed estimator to explore whether
or not the California Department of Transportation's decision to use
a reverse first price auction to award construction contracts is cost
minimizing. To do this, we use the method provided by Guerre, Perrigne,
and Vuong (2000) to calculate the firms' marginal costs using data
on their bids. We find that a kernel density estimator of these costs
does not satisfy Myerson's (1981) regularity condition everywhere;
however, the proposed density estimate, subject to the constraint
that Myerson's (1981) regularity condition is satisfied, appears to
follow the data closely. Our test also fails to reject that the population
density satisfies Myerson's (1981) regularity condition. 

In addition to the flexibility offered by the proposed framework,
there are three other advantages of the proposed method. First, the
notion of fidelity to the data that we optimize is independent of
the constraint, including the choice of $\rho$ in the case of $\rho-$concavity
constraints. Note that the objective function used in Koenker and
Mizera's (2010) approach, Rényi entropy of order $\rho,$ is dependent
on this constraint parameter. Also, adding a $\rho-$concavity constraint
to the maximum likelihood estimator would not provide a convincing
way to achieve this goal since this would not provide a convex optimization
problem for values of $\rho<0$ and the estimator does not exist when
$\rho<-1$ (Doss and Wellner, 2016). 

Second, the shape constraints of the estimator only binds in regions
in which it would not otherwise be satisfied. This is advantageous
when using shape constraints that are not sufficiently restrictive
to ensure parsimony by themselves. Moreover, the existence of the
unconstrained minimizer ensures that the density estimator exists,
regardless of the strength of the shape constraint. As discussed in
the preceding paragraph, this is not the case for the maximum likelihood
estimator. Although it is not our primary focus, we also provide an
option for relying on the shape constraint for parsimony. 

Third, the proposed algorithm solves an optimization problem over
a set of variables that grows sub-linearly in the sample size, so
the time complexity of the proposed algorithm compares favorably to
other shape constrained density estimators. 

The next section outlines the aspects of optimal transport that are
required to formulate our estimator. Galichon (2016) provides a more
comprehensive overview of the optimal transport literature, including
its many applications in economics. The third section defines the
estimator, provides the rate of convergence, and the asymptotic distribution
of the estimator. This section also provides a test for the null hypothesis
that the population density satisfies the shape constraint. The fourth
section proves that the optimization problem defining the estimator
is convex and provides an algorithm to calculate the estimator. Note
that initializing this algorithm at a reasonable approximation of
the density estimator provides a gain in computational efficiency,
and an algorithm for finding an approximation is provided in the appendix.
The sixth section generalizes the framework presented here to allow
for density estimation and inference subject to a much larger class
of shape constraints, with a focus on shape constraints that arise
in the economics literature. Specifically, this section provides sufficient
conditions for each of the results in the paper to hold under an arbitrary
set of shape constraints. This generalization is used in the seventh
section to provide evidence that the firms bidding on the California
Department of Transportation's construction contracts have marginal
cost distributions that satisfy Myerson's (1981) regularity condition. 

A few notational conventions will be useful in the subsequent sections.
For  $x\in\mathbb{R}^{m},$ we will denote the vector with an $i^{th}$
element defined by $\exp(x_{i})$ as $\exp(x),$ and a similar convention
will be used for $\log(x)$ and $x^{\rho}.$ Also, a diagonal matrix
with a diagonal equal to the vector $x$ will be denoted by $D_{x},$
an $m\times1$ vector of ones by \textbf{$\mathbf{1}_{m},$ }the identity
matrix by $I,$ element-wise division of the two vectors $x$ and
$y$ by $x\oslash y,$ element-wise multiplication by $x\otimes y,$
the Moore-Penrose pseudoinverse of the matrix $A$ as $A^{+},$ the
convolution between $f:\mathbb{R}^{d}\rightarrow\mathbb{R}^{1}$ and
$g:\mathbb{R}^{d}\rightarrow\mathbb{R}^{1}$ by $g(x)\ast f(x),$
the derivative of $f(x)$ with respect to $x$ by $\triangledown_{x}f(x),$
and a Dirac delta function centered at $z$ by $\delta_{z}(x).$ Also,
$sgn(x)$ will be used to denote a function that is $1$ when $x\geq0$
and $-1$ when $x<0.$ Since the proposed method requires discretizing
densities, say $\mu:\mathcal{A}\rightarrow\mathbb{R}^{1}$ for $\mathcal{A}\subset\mathbb{R}^{d},$
we will denote the points in the mesh as $\{\mathbf{a}_{i}\}_{i=1}^{m},$
 where $\mathbf{a}_{i}\in\mathbb{R}^{d}.$ We will also continue to
include parenthesis after functions, as in $\mu(x)$ or $\mu(\cdot),$
and exclude parenthesis when denoting $\mu\in\mathbb{R}^{m}$ with
elements $\mu_{i}=\mu(\mathbf{a}_{i}).$ 

\section{Optimal Transport }

Gaspard Monge formulated the theory of optimal transport in the $18^{th}$
century in order to derive the optimal method of moving a pile of
sand to a nearby hole of the same volume. Specifically, suppose that
both the pile of sand and the hole are defined on $\mathcal{A}\subset\mathbb{R}^{d},$
and we use the measures $\mathcal{M}_{0}:\mathcal{A}\rightarrow\mathbb{R}_{+}$
and $\mathcal{M}_{1}:\mathcal{A}\rightarrow\mathbb{R}_{+}$ to define
the volume of the pile and the hole respectively. Monge sought to
find a transportation plan, $T:\mathcal{A}\rightarrow\mathcal{A},$
that minimizes transportation costs while ensuring that the hole is
completely filled. 

Kantorovitch (1958) generalized this problem by describing the transportation
plan by the absolutely continuous measure $\psi:\mathcal{A}\times\mathcal{A}\rightarrow\mathbb{R}_{+}.$
For example, given $a_{1}\in\mathcal{A}$ and $a_{2}\in\mathcal{A},$
we can view the Radon-Nikodym derivative of $\psi(\cdot),$ $d\psi(a_{1},a_{2}),$
as the amount of mass moved from $a_{1}$ to $a_{2}$ under the transportation
plan, or \textit{coupling}, $\psi(\cdot).$ Feasibility of $\psi(\cdot)$
simply requires $\psi(a,\mathcal{A})=\mathcal{M}_{1}(a)$ and $\psi(\mathcal{A},a)=\mathcal{M}_{0}(a)$
for all $a\in\mathcal{A}.$ 

When $\mathcal{M}_{i}(\cdot)$ is absolutely continuous, there exists
$\mu_{i}(a)$ such that $\mathcal{M}_{i}(A)=\int_{A}\mu_{i}(a)da$
for each $A\subset\mathcal{A}$ by the Radon-Nikodym theorem. When
$\mathcal{M}_{i}(\cdot)$ also satisfies $\mathcal{M}_{i}(\mathcal{A})=1,$
then $\mu_{i}(a)$ is also a probability density function. Optimal
transport can be described without assuming $\mathcal{M}_{0}(\cdot)$
and $\mathcal{M}_{1}(\cdot)$ satisfy these conditions; however, since
our goal is density estimation, we will generally restrict our attention
to these cases for the rest of the paper. In addition we will define
(or constrain) all density functions to be continuous, with the exception
of $\delta_{z}(\cdot).$ We will use this notation to define $\Psi(\mu_{0}(\cdot),\mu_{1}(\cdot))$
as the set of feasible couplings. 

The most common cost function in optimal transport is simply squared
Euclidean distance. In this case the cost of moving one unit of earth
from $a_{1}\in\mathcal{A}$ to $a_{2}\in\mathcal{A}$ is proportional
to $\left\Vert a_{1}-a_{2}\right\Vert ^{2}.$ The resulting minimization
problem is then given by 

\begin{equation}
W_{0}(\mu_{0}(\cdot),\mu_{1}(\cdot)):=\underset{\psi\in\Psi(\mu_{0}(\cdot),\mu_{1}(\cdot))}{\min}\int_{\mathcal{A}\times\mathcal{A}}\left\Vert a_{1}-a_{2}\right\Vert ^{2}d\psi(a_{1},a_{2}),
\end{equation}

\noindent which we will refer to as the squared Wasserstein distance
(Mallows, 1972). $W_{0}(\mu_{0}(\cdot),$ $\mu_{1}(\cdot))$ has many
desirable properties, one being that $\sqrt{W_{0}(\mu_{0}(\cdot),\mu_{1}(\cdot))}$
satisfies all of the usual properties of a distance metric. $W_{0}(\mu_{0}(\cdot),\mu_{1}(\cdot))$
also metrizes weak convergence and convergence in the first two moments.
In other words, given a sequence of densities, $\{\mu_{0}(\cdot),\mu_{1}(\cdot),...,\mu_{n}(\cdot)\},$
we have $\lim_{i\rightarrow\infty}W_{0}(\mu_{0}(\cdot),\mu_{i}(\cdot))=0$
if and only if $\mathcal{M}_{i}$ converges weakly to $\mathcal{M}_{0}$
and the first two moments of $\mu_{i}$ converge to the first two
moments of $\mu_{0}.$ 

The Wasserstein distance between two distributions can be viewed as
a measure of distance over the domain of the densities rather than
in the direction of their range. For example, the Fréchet mean of
two Dirac delta functions, centered at $a$ and $b,$ in the spaces
of densities equipped with an $L^{2}$ norm is given by $\text{arg min}_{\nu(x)}\;\left\Vert \delta_{a}(x)-\nu(x)\right\Vert ^{2}+\left\Vert \delta_{b}(x)-\nu(x)\right\Vert ^{2}=\delta_{a}(x)/2+\delta_{b}(x)/2,$
while a similar notion of average in the spaces of densities equipped
with the Wasserstein distance is $\text{arg min}_{\nu(x)}\;W_{0}(\delta_{a}(x),\nu(x))+W_{0}(\delta_{b}(x),\nu(x))=\delta_{a/2+b/2}(x).$
To make this intuition more explicit, when $\mathcal{A}\subset\mathbb{R}^{1},$
one can show $W_{0}(\mu_{0}(\cdot),\mu_{1}(\cdot))$ can also be expressed
as $\int_{0}^{1}(Q_{0}(\tau)-Q_{1}(\tau))^{2}d\tau,$ where $Q_{0}(\tau)$
and $Q_{1}(\tau)$ are the quantile functions corresponding to $\mu_{0}(\cdot)$
and $\mu_{1}(\cdot)$ respectively. Thus, $(Q_{0}(\tau)-Q_{1}(\tau))^{2}$
represents a squared distance between two points in $\mathcal{A}$
(Villani, 2003). 

In practice augmenting the Wasserstein distance with a regularization
term ameliorates some numerical difficulties, which will be described
below in more detail. The regularized squared Wasserstein distance
is a generalization of $W_{0}(\mu_{0}(\cdot),\mu_{1}(\cdot)),$ and
is defined by

\begin{equation}
W_{\gamma}(\mu_{0}(\cdot),\mu_{1}(\cdot)):=\underset{\psi\in\Psi(\mu_{0}(\cdot),\mu_{i}(\cdot))}{\min}\int_{\mathcal{A}\times\mathcal{A}}\left\Vert a_{1}-a_{2}\right\Vert ^{2}d\psi(a_{1},a_{2})-\gamma H(\psi(\cdot)),
\end{equation}

\noindent where $\gamma\geq0$ and $H(\psi(\cdot)):=-\int_{\mathcal{A}\times\mathcal{A}}\log\psi(a_{1},a_{2})d\psi(a_{1},a_{2})$
is the Shannon entropy of $\psi(\cdot)$ (Cuturi, 2013; Cuturi and
Doucet, 2014). In the shape constrained density estimation setting,
the addition of this entropy term is advantageous for several reasons.
First, the objective function is strictly convex when $\gamma>0,$
so the optimal coupling will always be unique. Second, in practice
$\mathcal{A}$ must be discretized before finding the unregularized
Wasserstein distance, and the computational cost of solving for the
optimal coupling scales at least cubically in the number of points
in the mesh. Third, after discretizing, the minimizer of $W_{\gamma}(\mu_{0},\mu_{1})$
with respect to $\mu_{0}$ is often a more accurate representation
of the minimizer of $W_{0}(\mu_{0}(\cdot),\mu_{1}(\cdot)),$ when
$\gamma$ is set to a reasonably small value.\footnote{\noindent Minimizing $W_{0}(\mu_{0},\mu_{1})$ with respect to $\mu_{0}$
generally results in a minimizing density with many large discrete
changes. For more detail, see Figures 3.1, 3.2, and the accompanying
explanation in (Cuturi and Peyré, 2016). } Four, using $W_{\gamma}(\mu_{0}(\cdot),\mu_{1}(\cdot))$ allows us
to avoid assumptions in the next section regarding the existence of
the second moments of $\mu_{0}$ and $\mu_{1}.$ Lastly, we can find
the minimizer of (2) with a very computationally efficient algorithm
after discretizing, which we will describe next. 

To introduce the discretized counterparts of $d\psi(\cdot),\mu_{1}(\cdot),$
and $\mu_{0}(\cdot),$ recall our uniform mesh over $\mathcal{A}$
contains the vertices $\{\mathbf{a}_{i}\}_{i=1}^{m},$ and let $\mu_{0},\mu_{1}$
define $\mu_{0}(\mathbf{a}_{i}),\mu_{1}(\mathbf{a}_{i})$ respectively.
Also, let $M_{m\times m}$ so that $M_{ij}:=\left\Vert \mathbf{a}_{i}-\mathbf{a}_{j}\right\Vert ^{2},$
and $\psi_{m\times m}$ so that $\psi_{i,j}:=d\psi(\mathbf{a}_{i},\mathbf{a}_{j}).$
After discretizing, (2) can be written as 

\begin{equation}
W_{\gamma}(\mu_{0},\mu_{1}):=\underset{\psi}{\min}\;\sum_{i,j}\psi_{ij}M_{ij}+\gamma\psi_{ij}\log(\psi_{ij})\;\;\text{subject to:}
\end{equation}

\begin{equation}
\sum_{j}^{m}\psi_{ij}=\mu_{0i}\;\forall i\in\{1,2,...,m\}
\end{equation}
\begin{equation}
\sum_{i}^{m}\psi_{ij}=\mu_{1j}\;\forall i\in\{1,2,...,m\}.
\end{equation}

\noindent The corresponding Lagrangian is given by

\begin{equation}
\mathcal{L}=\left(\underset{i,j}{\sum}\gamma\psi_{ij}\log\left(\psi_{ij}\right)+\psi_{ij}M_{ij}\right)-\lambda_{0}^{T}\left(\sum_{j}^{m}\psi_{\cdot j}-\mu_{0}\right)-\lambda_{1}^{T}\left(\sum_{i}^{m}\psi_{i\cdot}-\mu_{1}\right),
\end{equation}

\noindent and the first order conditions imply
\begin{center}
$\psi_{ij}=\exp\left(\lambda_{0i}/\gamma-1/2\right)\exp\left(-M_{ij}/\gamma\right)\exp\left(\lambda_{1j}/\gamma-1/2\right).$
\par\end{center}

\noindent In other words, there exists $v,w\in\mathbb{R}_{+}^{m}$
such that the optimal coupling has elements $\psi_{ij}=K_{ij}w_{i}v_{j},$
where $K_{ij}:=\exp\left(-\left\Vert \mathbf{a}_{i}-\mathbf{a}_{j}\right\Vert ^{2}/\gamma\right),$
a symmetric $m\times m$ matrix. This can also be written as, 

\noindent 
\begin{equation}
\psi=D_{w}KD_{v},
\end{equation}

\noindent so adding the entropy term to the objective function reduces
the dimensionality of the optimization problem from $m^{2}$ to $2m.$
Sinkhorn (1967) shows that $\psi$ is unique. The iterative proportional
fitting procedure (IPFP) is an efficient method of computing these
vectors; see Krupp (1979). This method iteratively redefines $w$
so that $D_{w}Kv=\mu_{0},$ and subsequently redefines $v$ so that
$D_{v}Kw=\mu_{1},$ as summarized in Algorithm 1. Note that after
combining these equalities we have $D_{w}K(\mu_{1}\oslash(Kw))=\mu_{0},$
which will be used in subsequent sections. 
\begin{algorithm}[H]
\textbf{function}\texttt{ IPFP}$(K,\mu_{0},\mu_{1})$

$w\leftarrow\mathbf{1}_{m}$\textbf{ }

\textbf{until convergence: }

$\;\;\;\;$$v\leftarrow\mu_{1}\oslash(Kw)$

$\;\;\;\;$$w\leftarrow\mu_{0}\oslash(Kv)$ 

\textbf{return $w,v$}

\caption{The iterative proportional fitting procedure. }
\end{algorithm}

In the rest of the paper we will make substantial use of the dual
of (3)-(5). Cuturi and Doucet (2014) show that the dual is given by
the unconstrained optimization problem,

\begin{equation}
W_{\gamma}(\mu_{0},\mu_{1})=\underset{(x,y)\in\mathbb{R}^{2m}}{\max}\;x^{T}\mu_{0}+y^{T}\mu_{1}-\gamma\sum_{i,j}\exp\left((x_{i}+y_{j}-M_{ij})/\gamma\right).
\end{equation}

\noindent Note that $K=\exp\left(-M/\gamma\right),$ so the first
order conditions of (8) can be written as,

\begin{equation}
\mu_{0i}/\left(\sum_{j}\exp\left(y_{j}/\gamma\right)K_{ij}\right)=\exp\left(x_{i}/\gamma\right)
\end{equation}

\begin{equation}
\mu_{1j}/\left(\sum_{i}\exp\left(x_{i}/\gamma\right)K_{ij}\right)=\exp\left(y_{j}/\gamma\right).
\end{equation}

\noindent Note that after replacing $\exp\left(x/\gamma\right)$ with
$w$ and $\exp\left(y/\gamma\right)$ with $v,$ these formulas are
equivalent to the updates of $w$ and $v$ given in Algorithm 1. Also,
given $x$ and $y$ that satisfy (9)-(10), consider the vectors $\tilde{x}:=x-c$
and $\tilde{y}:=y+c,$ where $c\in\mathbb{R}.$ Since $\mu_{0}$ and
$\mu_{1}$ have the same sum, the objective function of (8), evaluated
at $\tilde{x}$ and $\tilde{y}$ must equal the objective function
evaluated at $x$ and $y.$ Since $\exp\left(\tilde{y}_{j}/\gamma\right)=\exp\left(y_{j}/\gamma\right)\exp\left(c/\gamma\right)$
and $\exp\left(\tilde{x}_{i}/\gamma\right)=\exp\left(x_{i}/\gamma\right)\exp\left(-c/\gamma\right),$
$\tilde{x}$ and $\tilde{y}$ must also satisfy (9)-(10). In other
words, while $v$ and $w$ are unique up to a multiplicative constant
on $w$ and one over this constant on $v,$ $y$ and $x$ are unique
up to the additive constant $c.$ 

A few comments regarding the effect of $\gamma$ on the optimal coupling
will also be useful in subsequent sections. Higher values of $\gamma$
correspond to placing a higher penalty on the negative entropy of
the coupling, so the optimal coupling becomes more dispersed as this
parameter is increased. Also, in the limit $\gamma\rightarrow0,$
$W_{\gamma}(\mu_{0},\mu_{1})$ converges to $W_{0}(\mu_{0},\mu_{1})$
at the rate $O(\exp(-1/\gamma))$ and $\psi$ converges to the optimal
unregularized coupling at this same rate; see Benamou et al. (2015)
and Cuturi (2013). In the next section will use $W_{\gamma}(\cdot)$
as an objective function to define the proposed estimator and show
how $\gamma$ impacts this estimator in more detail. 

\section{Shape-Constrained Density Estimation}

The input of the density estimator proposed in this paper is a kernel
density estimator, $\mu,$ based on $N$ i.i.d. datapoints, $\{z_{i}\}_{i=1}^{N},$
drawn from a uniformly continuous population density, $\mu^{\star}:\mathbb{R}^{d}\rightarrow\mathbb{R}_{+}^{1},$
with a bandwidth of $\sigma\geq0.$ Our results also hold when $\gamma$
and $\sigma$ are redefined to be functions of the form $c_{1}(\{z_{i}\}_{i})\gamma$
and $c_{2}(\{z_{i}\}_{i})\sigma,$ where $c_{j}(\{z_{i}\}_{i})\stackrel{p}{\rightarrow}c_{j}$
for $j\in\{1,2\}$ at any polynomial rate. In the interest of the
brevity of notation, we will only write the parameters in this way
when their randomness would have a non-trivial impact on the result.
$\gamma,\sigma$ and $m$ will also be dependent on $N,$ but we will
suppress this input throughout the paper and discuss our recommendations
for defining them after Theorem 2. 

With this notation in mind we can define the shape-constrained density
estimator as,
\begin{center}
$\underset{f}{\min}\;W_{\gamma}(f,\mu)\;\;\textnormal{subject to:}$
\par\end{center}

\begin{equation}
sgn(\rho)f^{\rho}\in\mathcal{K},
\end{equation}

\noindent where $\mathcal{K}$ is the cone of concave functions. Although
$\mathcal{K}$ is a convex set, the set of $\rho-$concave densities
is generally not convex. To ameliorate this problem, we will use a
similar formulation as Koenker and Mizera (2010) and solve for $g:=f^{\rho}.$
Note that no generality is lost in doing so, as there is a one to
one correspondence between $g$ and $f.$

For the clarity of the derivations in the next section, we will also
define $g$ to be a vector of length $m-1$ and refer to the element
of the vector $f,$ of length $m,$ that corresponds to this omitted
value as $f_{k}.$ We will set $f_{k}$ so that the density sums to
$m.$ In other words, the elements of the density $f$ that do not
correspond to this $k^{th}$ element, denoted by $f_{-k},$ will be
set equal to $g^{1/\rho}$ and $f_{k}$ will be set equal to $m-\sum_{i}g_{i}^{1/\rho}.$ 

Unlike optimizing over $g$ rather than $f,$ this can be seen as
a slight modification of our initial optimization problem, since we
will also exclude the shape constraints that depend on the $k^{th}$
element of $f.$ However, as discussed in the next section in more
detail, one can choose $k$ to correspond to an element on the boundary
of the mesh so that the estimator satisfies the shape constraint everywhere
on the interior of its domain. 

In a slight abuse of notation, we will also denote the objective function
as $W_{\gamma}(g^{1/\rho},\mu).$ In summary, we will define $W_{\gamma}(g^{1/\rho},$
$\mu)$ by,

\begin{equation}
\underset{(x,y)\in\mathbb{R}^{2m}}{\max}\;x_{-k}^{T}g^{1/\rho}+x_{k}\left(m-\sum_{i}g_{i}^{1/\rho}\right)+y^{T}\mu-\gamma\sum_{i,j}\exp\left((x_{i}+y_{j}-M_{ij})/\gamma\right),
\end{equation}

\noindent and the final form of our optimization problem is,

\begin{equation}
\underset{g}{\min}\;W_{\gamma}(g^{1/\rho},\mu)\;\;\textnormal{subject to:}
\end{equation}

\begin{equation}
g_{i}=\alpha_{i}+\beta_{i}\text{a}_{i},\;\;sgn(\rho)\left(\alpha_{i}-\alpha_{j}+(\beta_{i}-\beta_{j})^{T}\text{a}_{i}\right)\leq0\;\;\forall\;i,j\in\{2,...,m-1\},
\end{equation}

\noindent where $\alpha_{i}\in\mathbb{R}^{1},$ $\beta_{i}\in\mathbb{R}^{d},$
and $d$ is the dimension of the support of $\mu$ and $f.$ These
inequality constraints are used by Afriat (1972) to estimate production
functions with concavity constraints. They tend to provide a gain
in numerical accuracy relative to local concavity constraints. 

The following Lemma provides the limiting distribution of the estimator
after removing the shape constraints. This is achieved by showing
that the resulting density can be viewed as a kernel density estimator
with a bandwidth of $\sqrt{\sigma^{2}+\gamma/2}$ away from the edges
of the mesh over $\mathcal{A}.$ To avoid these boundary value effects,
we recommend enlarging the domain of $\mu$ and $\hat{f}$ to include
regions within approximately $3\sqrt{\sigma^{2}+\gamma/2}$ of the
datapoints. Alternatively, one could replace the matrix $K$ with
the direct application of a Gaussian filter. This approach has the
added benefit of reducing the computational complexity of Algorithm
1 to $O(m\log m),$ so this is our recommended approach when $d>2;$
see Solomon et al. (2015) for more details on this method. 

$\vphantom{c}$ 

\textbf{Lemma 1:}\textit{ Suppose $\mu$ is a kernel density estimate,
generated with a Gaussian kernel and a bandwidth $\sigma$ and $\mu^{\star}(\cdot)$
is uniformly continuous. Also, suppose $\sigma,$ $\gamma$ and $m$
are chosen so that $\sqrt{\sigma^{2}+\gamma/2}\stackrel{p}{\rightarrow}0,$
$N\sqrt{\sigma^{2}+\gamma/2}\stackrel{p}{\rightarrow}\infty,$ and
$\min_{i\neq j}\left\Vert {\bf a}_{i}-\mathbf{a}_{j}\right\Vert /\sqrt{\gamma}\rightarrow0$
as $N\rightarrow\infty.$  Then there exists $c\in\mathbb{R}^{1}$
so that the limiting density of $f_{unc}:=\text{arg min}_{f}\;W_{\gamma}(f,\mu)$
is given by,}\footnote{When $\mu^{\star}(\cdot)$ is differentiable at $\mathbf{a}_{i},$
\textit{$c=\triangle\mu_{i}^{\star}(x)/2\mid_{x=\mathbf{a}_{i}},$
}where $\triangle\mu_{i}^{\star}(x)$ denotes the Laplacian, $\sum_{j=1}^{d}\triangledown_{x_{j},x_{j}}\mu^{\star}(x).$
However, $c$ can be defined without assuming $\mu^{\star}(\cdot)$
is differentiable; Karunamuni and Mehra (1991) provide more details
on this approach. }
\begin{center}
$\sqrt{N(\sigma^{2}+\gamma/2)^{d/2}}\left(f_{unc,i}-\mu_{i}^{\star}+c(\sigma^{2}+\gamma/2)^{d}\right)\stackrel{d}{\rightarrow}N\left(0,\mu_{i}^{\star}/\left(2\sqrt{\pi}\right)^{d}\right)$
\par\end{center}

\textit{Proof: }To find \textit{$f_{unc},$} consider the optimization
problem, 

\begin{equation}
\underset{\psi}{\min}\;\sum_{i,j}\psi_{ij}M_{ij}+\gamma\psi_{ij}\log(\psi_{ij})\;\;\text{subject to:}
\end{equation}

\begin{equation}
\sum_{i}^{m}\psi_{ij}=\mu_{j}\;\forall j\in\{1,2,...,n\}
\end{equation}

\noindent The corresponding Lagrangian is 
\begin{center}
$\mathcal{L}=\left(\sum_{i,j}\gamma\psi_{ij}\log(\psi_{ij})+\psi_{ij}M_{ij}\right)+\lambda_{0}^{T}\left(\sum_{i}\psi_{i\cdot}-\mu\right),$
\par\end{center}

\noindent and the first order conditions imply that there exists $v\in\mathbb{R}^{m}$
such that

\noindent 
\begin{equation}
\psi=KD_{v}.
\end{equation}

\noindent Note that convexity of negative entropy implies that the
optimal coupling will correspond to a minimizer. After combining this
equality with the constraints, we have

\begin{equation}
\sum_{i}^{m}\psi_{ij}=v_{j}\sum_{i}^{m}K_{ij}=\mu_{j},
\end{equation}

\noindent which implies 
\begin{center}
$v_{j}=\mu_{j}/\left(\sum_{i}^{m}K_{ij}\right).$
\par\end{center}

\noindent Let $\kappa:=K\mathbf{1}.$ Now we can find $f_{unc}$ by
finding the other marginal of $\psi,$ 
\begin{center}
$f_{unc}:=K\left(\mu\oslash\kappa\right).$
\par\end{center}

\noindent Let $\phi_{\eta}:\mathbb{R}^{d}\rightarrow\mathbb{R}^{1}$
be a Gaussian density with variance $\eta I_{d}$ and mean $\mathbf{0}_{d}.$
Suppose $\nu:\mathbb{R}^{d}\rightarrow\mathbb{R}^{1}$ is a continuous
function. Then, 
\begin{center}
$K\nu(\mathbf{a})=\sum_{i=1}^{m}\exp(-\left\Vert \mathbf{a}_{i}-\mathbf{a}_{j}\right\Vert /\gamma)\nu(\mathbf{a}_{i})$
\par\end{center}

\noindent \begin{center}
$\approx\frac{m\sqrt{\pi}}{\gamma}\sum_{i=1}^{m}\phi_{\gamma/2}\left(\mathbf{a}_{i}-\mathbf{a}_{j}\right)\nu(\mathbf{a}_{i})$
\par\end{center}

\noindent by the definition of $K,$ so $K$ can be viewed as a linear
operator that discretizes the convolution $\nu(\cdot)\ast\phi(\cdot)\sqrt{\pi}/\gamma.$
Since the distance between adjacent points decreases at a rate that
is faster than $\sqrt{\gamma/2},$ and the convex hull of the data
is a strict subset of the convex hull of $\{\mathbf{a}_{i}\}_{i},$
the error of this approximation converges to zero on the convex hull
of the data. Thus, as $N$ diverges we have $\kappa_{i}\rightarrow\sqrt{\pi}/(m\gamma)$
for all $i$ in the convex hull of the data. 

Since $\mu$ is a kernel density estimator, it can be expressed as
$\mu=\left(\sum_{i}\delta_{z_{i}}(\cdot)\ast\phi_{\sigma^{2}}/N\right),$
$(\mathbf{a})$ which implies that $f_{unc}$ converges to $\left(\sum_{i}\delta_{z_{i}}(\cdot)\ast\phi_{\sigma^{2}}\ast\phi_{\gamma/2}/N\right)(\mathbf{a}).$
The convolution of two Gaussian densities is $\phi_{\sigma^{2}}\ast\phi_{\gamma/2}=\phi_{\sigma^{2}+\gamma/2},$
so $f_{unc}=\left(\sum_{i}\delta(z_{i})\right)\ast\phi_{\sigma^{2}+\gamma/2}(y)/N,$
which defines a kernel density estimator with a bandwidth of $\sqrt{\sigma^{2}+\gamma/2},$
and Parzen (1962) provides the limiting density of the kernel density
estimator when $\sqrt{\sigma^{2}+\gamma/2}\rightarrow0$ and $N\sqrt{\sigma^{2}+\gamma/2}\rightarrow\infty.$ 
\begin{flushright}
$\square$
\par\end{flushright}

The following theorem provides the limiting density of the estimator
with the primary additional assumption that $(\mu^{\star}(x))^{\rho}$
is strictly concave. Afterward, we will move onto results that relax
this assumption. 

$\vphantom{c}$ 

\textbf{Theorem 2:}\textit{ Suppose the assumptions from Lemma 1 hold.
If $\mu^{\star}$ is in the interior of the feasible set, $N\sqrt{c_{2}(\{z_{i}\}_{i})\sigma^{2}+c_{1}(\{z_{i}\}_{i})\gamma/2}/\log(N)\rightarrow\infty,$
and for $j\in\{1,2\},$ $c_{j}(\{z_{i}\}_{i})$ converges in probability
to a constant, then there exists $c\in\mathbb{R}^{1}$ so that,}
\begin{center}
$\sqrt{N(\sigma^{2}+\gamma/2)^{d/2}}\left(\hat{f}_{i}-\mu_{i}^{\star}+c(\sigma^{2}+\gamma/2)^{d}\right)\stackrel{d}{\rightarrow}N\left(0,\mu_{i}^{\star}/\left(2\sqrt{\pi}\right)^{d}\right).$
\par\end{center}

\textit{Proof:} Since $W_{\gamma}(\cdot)$ is differentiable, we can
apply the envelope theorem to the dual problem (8) to show that $\triangledown_{f}W_{\gamma}(f,\mu)=x.$
This implies that the $\tilde{f}\in\mathbb{R}^{m}$ is a critical
point of $f\mapsto W_{\gamma}(f,\mu)$ if and only if it has a corresponding
dual variable in (8) of $x=\mathbf{0},$ so $f_{unc}$ is the unique
minimizer of $W_{\gamma}(\cdot)$ by the proof of Lemma 1 and the
definition $w:=\exp(x/\gamma).$\footnote{Uniqueness of $f_{unc}$ can also be shown using strict convexity
of $W_{\gamma}(f,\mu)$ in $f,$ which will be shown in Theorem 6. } This implies $\hat{f}=f_{unc}$ when $f_{unc}$ is feasible. Einmahl
and Mason (2005) show that $f_{unc}\stackrel{a.s.}{\rightarrow}\mu^{\star}$
under the additional assumptions of the theorem when using data dependent
bandwidths, as in the statement of the theorem. The result follows
from the fact that $\mu^{\star}$ is in the interior of the feasible
set. 
\begin{flushright}
$\square$
\par\end{flushright}

\textbf{Remark 1:} Algorithm 1 can be slow to converge when $\gamma$
is chosen to be too small, but our assumption that $\gamma\rightarrow0$
as $N\rightarrow\infty$ is not problematic when the assumptions of
the theorem hold. To see this note that $f_{unc}$ can be written
as $K(\mu\oslash(K\mathbf{1})).$ Evaluating Algorithm 1 at input
densities $\mu$ and $f_{unc}$ would result in the algorithm first
initializing $w$ as $\mathbf{1}_{m}$ and then define $v$ to be
$\mu\oslash(K\mathbf{1}).$ The $w-$update would redefine $w$ to
be $f_{unc}\oslash K(\mu\oslash(K\mathbf{1})),$ but since this is
also equal to $\mathbf{1},$ the algorithm has already converged.
For input densities $f$ and $\mu,$ it is also generally the case
that Algorithm 1 tends to converge at a faster rate for densities
that are closer to $f_{unc}.$ 
\begin{flushright}
$\square$
\par\end{flushright}

\textbf{Remark 2:} Our more general convergence result (Theorem 4)
provides the same rate of convergence as Theorem 2 without requiring
that $\mu^{\star}$ is an interior point of the feasible set, so we
can compare this rate with the corresponding rate of convergence of
shape constrained maximum likelihood estimators. Seregin and Wellner
(2010) show that the pointwise minimax absolute error loss of the
$\log-$concave constrained maximum likelihood estimators at $x\in\mathcal{A}$
is $N^{-2/(d+4)}$ when $\mu^{\star}(x)$ is twice differentiable,
$x$ is in the interior of the domain of $\mu^{\star}(\cdot),$ and
the Hessian has full rank. Theorem 2 implies that our estimator also
obtains these bounds when $\sqrt{\sigma^{2}+\gamma/2}$ is chosen
so that it converges to zero at the optimal rate, in the sense of
minimizing mean squared error, of $O_{p}(N^{-1/(d+4)}).$ 
\begin{flushright}
$\square$
\par\end{flushright}

In practice, $\sqrt{\sigma^{2}+\gamma/2}$ has a more noticeable impact
on the resulting density estimator than the individual values of $\sigma$
and $\gamma,$ so we recommend fixing $\gamma/\sigma^{2}$ to be a
constant and focusing on the choice of $\sqrt{\sigma^{2}+\gamma/2}.$
Since increasing $\gamma/\sigma^{2}$ tends to result in an increase
in the rate of convergence and numerical stability of Algorithm 1,
our recommendation is $\gamma/\sigma^{2}=8,$ which was used in all
of the applications and examples given below. 

From a numerical standpoint, it is possible to set $\sqrt{\sigma^{2}+\gamma/2}$
so that the smoothing provided by $\gamma$ and $\sigma$ is negligible
and parsimony is almost entirely ensured by the shape constraints.
Since the stability of Algorithm 1 is the only limiting factor on
how small we can make $\sqrt{\sigma^{2}+\gamma/2},$ we can use this
fact to estimate a lower bound on the value $\sqrt{\sigma^{2}+\gamma/2}.$
Specifically, we can estimate the lowest possible value of $\sqrt{\sigma^{2}+\gamma/2}$
that still results in convergence of Algorithm 1 to within a given
tolerance in a fixed number of iterations, say 100, which can be achieved
using a root finding algorithm. This requires an approximation of
the density estimate, and Appendix B provides a method for finding
an approximation in a computationally efficient manner. When running
the main optimization algorithm described in the next section, we
recommend increasing $\gamma$ after this root is found by approximately
one fourth and increasing the number of iterations used in Algorithm
1 by a factor of approximately ten to ensure the accuracy of the Hessian. 

In our tests this approach resulted in density estimates that were
surprisingly similar to estimates using the method described by Koenker
and Mizera (2010). However, it is likely that the condition \textit{$N\sqrt{\sigma^{2}+\gamma/2}\rightarrow\infty$}
would not hold in this case, which is a requirement of Theorem 2.
If $N\sqrt{\sigma^{2}+\gamma/2}\rightarrow0$ and we redefine the
location of the vertices in the mesh, $\mathbf{a},$ to be equal to
the datapoints, then asymptotically $\mu_{i}$ and $f_{unc,i}$ would
only be influenced by a single datapoint. The following theorem shows
that we can at least ensure $\hat{f}\overset{p}{\rightarrow}\mu^{\star}$
in the case $\gamma=\sigma=0.$

$\vphantom{c}$ 

\textbf{Theorem 3:}\textit{ Suppose the locations of the vertices
in the mesh $\mathbf{a}$ are defined to be the datapoints, $\mu^{\star}(z)>0$
on its bounded domain $\mathcal{A},$ and $\mu_{i}=1/N$ for $i\in\{1,2,...,N\}.$
If there exists $\epsilon>0$ such that $\int_{\mathcal{A}}\left|x\right|^{2+\epsilon}d\mu^{\star}<\infty,$
$\mu^{\star}$ is $\rho-$concave, and }$\gamma=\sigma=0,$\textit{
then }$\hat{f}\overset{p}{\rightarrow}\mu^{\star}.$

\textit{Proof:} The Wasserstein distance between density functions,
denoted by $W_{0}(\nu_{1}(z),$ $\nu_{2}(z)),$ is commonly approximated
by discretizing, using $\nu_{j,i}=\nu_{j}(z_{i})$ for $j\in\{1,2\}$
and $z_{i}\in\{\mathbf{a}_{i}\}_{i},$ and solving a linear programming
problem, as discussed in the previous section. As long as $\min\{r\mid\mathcal{A}\subset\cup_{i=1}^{m}B_{r}(\mathbf{a}_{i})\}$
goes to zero asymptotically, $W_{0}(\nu_{1},\nu_{2})$ converges to
its non-discretized counterpart, $W_{0}(\nu_{1}(z),\nu_{2}(z));$
see for example, (Cuturi and Peyré, 2016). Since \textit{$\mu^{\star}(z)>0$
}for all $z\in\mathcal{A}$ and $\{\mathbf{a}_{i}\}_{i}=\{z_{i}\}_{i},$
this condition holds asymptotically, so showing $W_{0}(\hat{f}(z),\mu^{\star}(z))\stackrel{p}{\rightarrow}0,$
where $\hat{f}(z)$ is the linear interpolation of $\{(\mathbf{a}_{i},\hat{f}_{i})\}_{i=1}^{N},$
will imply the result. 

Let $\mu(z)=\sum_{i}\delta_{z_{i}}(z)/N$ and $f(z)$ be an arbitrary
proper density with \textit{$\int_{\mathcal{A}}\left|x\right|^{2+\epsilon}df.$
}After using the triangle inequality twice, we can bound $W_{0}(f(\cdot),\mu(\cdot))$
by $W_{0}(f(\cdot),\mu^{\star}(\cdot))$ $\pm W_{0}(\mu(\cdot),\mu^{\star}(\cdot))$
for all $N>1.$ Note that Wasserstein distances metrize weak convergence,
in the sense that $W_{0}(\nu_{0}(z),\nu_{N}(z))\stackrel{p}{\rightarrow}0$
if and only if the distribution corresponding to $\nu_{N}(z)$ converges
weakly to that of $\nu_{0}(z)$ and the first two moments of $\nu_{N}(z)$
converge to the first two moments of $\nu_{0}(z);$ see for example,
Theorem 7.12 of (Villani, 2003). Thus, $W_{0}(\mu(\cdot),\mu^{\star}(\cdot))\stackrel{p}{\rightarrow}0$
by the result of Kiefer and Wolfowitz (1956), so our bounds on $W_{0}(f(\cdot),\mu(\cdot))$
imply that $W_{0}(f(\cdot),\mu(\cdot))\stackrel{p}{\rightarrow}W_{0}(f(\cdot),\mu^{\star}(\cdot))$
for all such $f(\cdot).$ 

Note that $W_{0}(f(\cdot),\mu^{\star}(\cdot))$ obtains its minimum
value with respect to $f(\cdot)$ at densities with distribution functions
that are arbitrarily close to the distribution function of $\mu^{\star}(\cdot).$
Since $\mu^{\star}$ is $\rho-$concave by definition, and $\hat{f}(\cdot)$
is also by our constraint, this distribution has well defined and
continuous density function by Aleksandrov's theorem. Thus, $\hat{f}(\cdot)\overset{p}{\rightarrow}\mu^{\star}(\cdot)$
by the standard argument for consistency of an M-estimator; see for
example, Theorem 5.7 of (van der Vaart, 2000).
\begin{flushright}
$\square$
\par\end{flushright}

For the five reasons discussed in the previous section, our primary
focus is on cases in which $\mu,$ and $f_{unc},$ are also consistent
estimators, so we will move onto our recommended approach for setting
$\sqrt{\sigma^{2}+\gamma/2}$ and $m.$ Like kernel density estimators,
the mean squared error of $f_{unc}$ is minimized when $\sqrt{\sigma^{2}+\gamma/2}$
is $O(N^{-1/(d+4)}).$ Also, any of the standard techniques for setting
the bandwidth of kernel density estimators are reasonable methods
of setting $\sqrt{\sigma^{2}+\gamma/2},$ including cross validation
or a rule-of-thumb bandwidth estimator; for examples of rule-of-thumb
bandwidth estimators see (Silverman, 1986; Scott, 1992). Since the
shape constraint also helps to ensure the parsimony of the density
estimate, these rules should generally be regarded as upper bounds
on $\sqrt{\sigma^{2}+\gamma/2}.$ In practice, using Scott's (1992)
rule-of-thumb multiplied by $2/3$ works well. After dividing each
dimension of the dataset, $\{z_{i,j}\}_{i}$ for $j\in\{1,2,...,d\},$
by $\min(\text{IQR}(\{z_{i,j}\}_{i})/1.349,\hat{\sigma}(\{z_{i,j}\}_{i})),$
combining this with $\gamma/\sigma^{2}=8,$ results in $\sigma\approx N^{-1/(d+4)}/3$
and $\gamma\approx N^{-2/(d+4)}4/5.$ 

The choice of $m$ is likely less critical than that of $\sqrt{\sigma^{2}+\gamma/2}.$
The effect of $m$ on the estimator can be viewed as analogous to
the effect of the size of a Gaussian filter on its output. In practice,
these filters are often constructed by discretizing a Gaussian density
over a mesh with a resolution equal to the standard deviation. Using
this as a lower bound in our setting yields, $m\sqrt{\gamma/2}/d\geq1.$
The gain in accuracy from increasing $m$ beyond the point $m\sqrt{\gamma/2}/d=1$
generally diminishes fairly quickly when $m$ is set to larger values,
so we recommend choosing $m$ so that $m=\left\lceil N^{\beta}d/\sqrt{\gamma/2}\right\rceil ,$
where $\left\lceil \cdot\right\rceil $ is the ceiling function, for
any $\beta>0.$ In the interest of specificity, we recommend using
$m=\left\lceil N^{1/5}d/\sqrt{\gamma/2}\right\rceil .$ 

If the set of constraints that bind asymptotically is known, it is
straightforward to find the asymptotic distribution of $\hat{f}$
when $\mu^{\star}$ is not $\rho-$concave; however, this is rarely
the case in practice. For this reason, finding confidence intervals
for estimators subject to shape constraints is an active area of research
in nonparametric econometrics. The difficulties in these cases are
due to the estimators, when viewed as functions of the data, not being
sufficiently smooth (either differentiable or Hadamard differentiable)
at points in which an inequality constraint goes from slack to active,
and these notions of smoothness are requirements of many of the commonly
used methods for deriving limiting distributions (Andrews, 2000). 

The following theorem establishes the limiting distribution of the
estimator conditional on the set of constraints that bind asymptotically.
Note that the set of active constraints converges when $\sqrt{\sigma^{2}+\gamma/2}$
converges to zero at its optimal rate. Thus, the limiting distribution
part of the proof might be useful with a rather large sample size.
However, since the set of constraints that binds in a finite sample
may not be equal to the set of constraints that bind asymptotically,
in many cases the primary value of this result is that it provides
the limiting value of the estimator without assuming that $(\mu^{\star})^{\rho}$
is strictly concave. 

There are several possible methods to avoid conditioning on the set
of active constraints, which is an area of ongoing research. One interesting
possibility would be to adopt a similar method as Horowitz and Lee
(2017) to the present setting. Their technique involves explicitly
defining the set of active constraints as those constraints that would
bind otherwise or that would ``nearly bind.'' Asymptotically correct
coverage follows from consistency of this conservative estimate of
the active set. 

$\vphantom{c}$ 

\textbf{Theorem 4:}\textit{ Suppose the assumptions from Lemma 1 hold.
In addition, suppose $\sigma N\rightarrow\infty.$ Then, conditional
on the set of constraints that are active asymptotically, say $\Omega,$
there exists $G,$ as defined in (23-25), and $c\in\mathbb{R}^{m}$
such that}
\noindent \begin{center}
\textit{$\sqrt{N(\sigma^{2}+\gamma/2)^{d/2}}\left(\hat{f}-\tilde{\mu}+D_{c}(\sigma^{2}+\gamma/2)^{d}\right)\stackrel{d}{\rightarrow}N\left(0,G^{T}D_{\mu^{\star}/(2\sqrt{\pi})^{d}}G\right),$ }
\par\end{center}

\noindent \textit{where $\tilde{\mu}$ is the Wasserstein projection
of $\mu^{\star}$ onto the set of feasible densities. Specifically,
if $\tilde{g}$ is the minimizer of}
\begin{center}
$\underset{g}{\min}\;W_{\gamma}(g^{1/\rho},\mu^{\star})\;\;\textnormal{subject to:}$
\par\end{center}

\begin{center}
$sgn(\rho)g\in\mathcal{K},$
\par\end{center}

\noindent \textit{then $\tilde{\mu}_{-k}:=\tilde{g}^{1/\rho}$ and
$\tilde{\mu}_{k}=m-\sum_{i}\tilde{g}_{i}^{1/\rho}$ in the case of
a $\rho-$concavity constraint. }

\textit{Proof:} We only consider the case in which $d=1$ and $k=m$
for the sake of clarity, but generalizing the proof to higher dimensions
is straightforward. Since $d=1,$ we will assume all sets containing
the numerical labels of vertices in the mesh are ordered sequentially.
For example, $\Omega_{i}$ will be viewed as the $i^{th}$ largest
vertex label in which the constraint binds. Let $\dot{\Omega}:=\{i\mid i\notin\Omega\land i\pm1\in\Omega\},$
the boundary of the complement of $\Omega.$ 

Since $\sqrt{\sigma^{2}+\gamma/2}\rightarrow0,$ we have $\sigma\rightarrow0.$\textit{
}Since this is the case, and since \textit{$\sigma N\rightarrow\infty$
}and $\mu^{\star}$ is continuous, $\mu$ converges uniformly to $\mu^{\star}$
(Parzen, 1962). We can view $\hat{f}$ as a continuous function of
$\mu,$ so the continuous mapping theorem implies $\hat{f}$ converges
to $\tilde{\mu}.$ 

Let the matrix $A_{\left|\Omega\right|\times(\left|\Omega\right|+\left|\dot{\Omega}\right|)}$
be defined so that $Ag_{\Omega\cup\dot{\Omega}}$ is the second difference
of $g_{\Omega\cup\dot{\Omega}},$ so the binding constraints can be
denoted by $Ag_{\Omega\cup\dot{\Omega}}\leq\mathbf{0}.$\footnote{In the interest of the simplicity of the exposition, we are defining
$A$ using local concavity constraints rather than Afriat's (1972)
formulation of concavity constraints. The nonzero elements of each
row of $A$ are given by $(A_{i,j-1},A_{i,j},A_{i,j+1})=sgn(\rho)(1,-2,1).$
Note that $j>1,$ and, since the sets $\Omega$ and $\Omega\cup\dot{\Omega}$
are ordered by the vertex labels, this matrix is in row echelon form
by construction. Thus, $A$ has full rank, so $AA^{T}$ is nonsingular. } Since we assumed that $\Omega$ is known, $\hat{g},$ the vector
of Lagrange multipliers, $\lambda,$ and $x$ are defined by the following
system of equations,

\begin{equation}
(x_{\Omega\cup\dot{\Omega}}-x_{k})\otimes\hat{g}_{\Omega\cup\dot{\Omega}}^{1/\rho-1}/\rho=-A^{T}\lambda
\end{equation}

\begin{equation}
(x_{-\Omega\cap-\dot{\Omega}}-x_{k})\otimes\hat{g}_{-\Omega\cap-\dot{\Omega}}^{1/\rho-1}/\rho=\mathbf{0}
\end{equation}

\begin{equation}
A\hat{g}_{\Omega}=\mathbf{0}
\end{equation}

\begin{equation}
[\begin{array}{cc}
\left(\hat{g}^{1/\rho}\right)^{T} & m-\sum_{i}\hat{g}_{i}^{1/\rho}\end{array}]^{T}=\exp(x/\gamma)\otimes\left(K\left(\mu\oslash(K\exp(x/\gamma))\right)\right),
\end{equation}

\noindent where the final equality results from combining the first
order conditions of the optimization problem defining $W_{\gamma}(g^{1/\rho},\mu),$
which are given by (9) and (10). 

Theorem 5 will establish that $W_{\gamma}(g^{1/\rho},\mu)$ is strictly
convex in $g.$ Since this is the case, the solution of this system
is unique. The implicit function theorem, applied to (19)-(22), implies
that we can view $\hat{g},$ and thus also $\hat{f},$ as a differentiable
function of $\mu.$ After simplifying this system, we will find this
limiting density using the delta method.

We will begin by deriving $\triangledown_{\mu}x$ using (22). Recall
$w:=\exp(x/\gamma),$ and let $h_{1}(x,\mu)$ be defined as,
\begin{center}
$h_{1}(x,\mu):=\hat{f}-D_{w}K\left(\mu\oslash(Kw)\right).$
\par\end{center}

\noindent (22) can be written as $h_{1}(x,\mu)=\mathbf{0}.$ Recall
the following equalities from the previous section: $\psi=D_{w}KD_{v},$
$f=D_{w}Kv,$ and $\mu=D_{v}Kw.$ These imply
\begin{center}
$\triangledown_{\mu}h_{1}(x,\mu)=-D_{w}KD_{\mathbf{1}\oslash(Kw)}$
\par\end{center}

\begin{center}
$=-D_{w}KD_{v\oslash\mu}=-\psi D_{\mathbf{1}\oslash\mu},$
\par\end{center}

\noindent and 
\begin{center}
$\triangledown_{w}h_{1}(x,\mu)=D_{w}KD_{\mu\oslash(Kw)^{2}}K-D_{K(\mu\oslash(Kw))}$ 
\par\end{center}

\begin{center}
$=D_{w}KD_{v}D_{\mathbf{1}\oslash\mu}D_{v}K-D_{\hat{f}\oslash w}$
\par\end{center}

\begin{center}
$=\left(\psi D_{\mathbf{1}\oslash\mu}\psi^{T}-D_{\hat{f}}\right)D_{\mathbf{1}\oslash w}.$
\par\end{center}

\noindent The implicit function theorem implies,
\begin{center}
$\triangledown_{\mu}w=D_{w}\left(\psi D_{\mathbf{1}\oslash\mu}\psi^{T}-D_{\hat{f}}\right)^{+}\psi D_{\mathbf{1}\oslash\mu}.$
\par\end{center}

\noindent Since $x=\gamma\log(w),$ we have, 
\begin{center}
$\triangledown_{\mu}x=\gamma\left(\psi D_{\mathbf{1}\oslash\mu}\psi^{T}-D_{\hat{f}}\right)^{+}\psi D_{\mathbf{1}\oslash\mu}.$
\par\end{center}

(21) implies that each element in $\hat{g}_{\Omega}$ can be expressed
as a mean of its neighbors, so we can express all of the elements
of $\hat{g}_{\Omega}$ as a weighted mean of $\hat{g}_{\dot{\Omega}}.$
In other words, there exists $C$ such that $\hat{g}_{\Omega\cup\dot{\Omega}}=C\hat{g}_{\dot{\Omega}}.$
(19) implies that, given $x$ and $g_{\dot{\Omega}},$ $\lambda$
is given by,
\begin{center}
$-\left(AA^{T}\right)^{-1}AD_{x_{\Omega\cup\dot{\Omega}}-x_{k}}(C\hat{g}_{\dot{\Omega}})^{1/\rho-1}/\rho=\lambda.$
\par\end{center}

\noindent We will use the additional $\left|\Theta\right|$ equations
in (19) to define $\hat{f}_{\dot{\Omega}}.$ After replacing each
instance of $\hat{g}_{\dot{\Omega}}$ with $\hat{f}_{\dot{\Omega}}^{\rho}$
and using this definition of $\lambda,$ we can write (19) as $h_{2}(x,\hat{f}_{\dot{\Omega}})=\mathbf{0},$
where $h_{2}(x,\hat{f}_{\dot{\Omega}})$ is defined as 
\begin{center}
$(x_{\dot{\Omega}}-x_{k})\otimes\hat{f}_{\dot{\Omega}}^{1-\rho}+\tilde{A}\left((x_{\Omega\cup\dot{\Omega}}-x_{k})\otimes(C\hat{f}_{\dot{\Omega}}^{\rho})^{1/\rho-1}\right),$
\par\end{center}

\noindent and $\tilde{A}:=A_{\cdot,\dot{\Omega}}^{T}\left(AA^{T}\right)^{-1}A.$
This implies 
\begin{center}
$\triangledown_{\hat{f}_{\dot{\Omega}}}h_{2}(\cdot)=(1-\rho)\left(D_{(x_{\dot{\Omega}}-x_{k})}\hat{f}_{\dot{\Omega}}^{-\rho}+\tilde{A}D_{x_{\Omega\cup\dot{\Omega}}-x_{k}}D_{(C\hat{f}_{\dot{\Omega}}^{\rho})^{1/\rho-2}}CD_{\hat{f}_{\dot{\Omega}}^{\rho-1}}\right),$
\par\end{center}

\noindent and
\begin{center}
$\triangledown_{x_{\Omega\cup\dot{\Omega}}-x_{k}}h_{2}(\cdot)=B+\tilde{A}D_{(C\hat{f}_{\dot{\Omega}}^{\rho})^{1/\rho-1}},$
\par\end{center}

\noindent where $B_{\left|\dot{\Omega}\right|\times\left|\Omega\cup\dot{\Omega}\right|}$
is defined so that $B_{i,j}$ is equal to $\hat{f}_{\dot{\Omega}}^{1-\rho}$
when $i,j$ satisfies $\dot{\Omega}_{i}=\{\Omega\cup\dot{\Omega}\}_{j}$
and zero otherwise. The implicit function theorem implies

$\triangledown_{x_{\Omega\cup\dot{\Omega}}-x_{k}}\hat{f}_{\dot{\Omega}}=-\left(D_{(x_{\dot{\Omega}}-x_{k})}\hat{f}_{\dot{\Omega}}^{-\rho}+\tilde{A}D_{(x_{\Omega\cup\dot{\Omega}}-x_{k})\otimes(C\hat{f}_{\dot{\Omega}}^{\rho})^{1/\rho-2}}CD_{\hat{f}_{\dot{\Omega}}^{\rho-1}}\right)^{-1}$
\begin{center}
$\cdot\left(B+\tilde{A}D_{(C\hat{f}_{\dot{\Omega}}^{\rho})^{1/\rho-1}}\right)/(1-\rho),$
\par\end{center}

\noindent so 
\begin{center}
$G_{\dot{\Omega},\cdot}:=\triangledown_{\mu}\hat{f}_{\dot{\Omega}}=\triangledown_{x_{\Omega\cup\dot{\Omega}}-x_{k}}\hat{f}_{\dot{\Omega}}P\triangledown_{\mu}x=$
\par\end{center}

$-\left(D_{(x_{\dot{\Omega}}-x_{k})}\hat{f}_{\dot{\Omega}}^{-\rho}+\tilde{A}D_{(x_{\Omega\cup\dot{\Omega}}-x_{k})\otimes(C\hat{f}_{\dot{\Omega}}^{\rho})^{1/\rho-2}}CD_{\hat{f}_{\dot{\Omega}}^{\rho-1}}\right)^{-1}$

\begin{equation}
\cdot\left(B+\tilde{A}D_{(C\hat{f}_{\dot{\Omega}}^{\rho})^{1/\rho-1}}\right)P\left(\psi D_{\mathbf{1}\oslash\mu}\psi^{T}-D_{\hat{f}}\right)^{+}\psi D_{\mathbf{1}\oslash\mu}\gamma/(1-\rho),
\end{equation}

\noindent where $P_{\left|\Omega\cup\dot{\Omega}\right|\times m}:=\triangledown_{x}(x_{\Omega\cup\dot{\Omega}}-x_{k})$
is defined so that $P_{i,j}$ is equal to $1$ when $i,j$ satisfies
$\{\Omega\cup\dot{\Omega}\}_{i}=j,$ $-1$ when $j=m,$ and zero otherwise.

Note that (20) implies $x_{i}=x_{k}$ for all $i\in\{j\mid j\notin\Omega\cap j\notin\dot{\Omega}\}.$
The proof of Theorem 2 shows that this condition defines $f_{unc,i},$
so we have $f_{unc,i}=\hat{f}_{i}$ for all such $i.$ This implies, 

\begin{equation}
G_{-\Omega\cap-\dot{\Omega},\cdot}:=\triangledown_{\mu}\hat{f}_{-\Omega\cap-\dot{\Omega}}=K_{-\Omega\cap-\dot{\Omega},\cdot}.
\end{equation}

Lastly, writing the equations defining $\hat{f}_{\Omega}$ in terms
of $\hat{f}_{\dot{\Omega}}$ yields $\hat{f}_{\Omega}=(C\hat{f}_{\dot{\Omega}}^{\rho})^{1/\rho},$
so we have,

\begin{equation}
G_{\Omega,\cdot}:=\triangledown_{\mu}\hat{f}_{\Omega}=D_{(C\hat{f}_{\dot{\Omega}}^{\rho})^{1/\rho-1}}CD_{\hat{f}_{\dot{\Omega}}^{\rho-1}}G_{\dot{\Omega},\cdot}.
\end{equation}

\begin{flushright}
$\square$
\par\end{flushright}

There are also a few options for testing if a population density satisfies
a shape constraint. Hypothetically, one could consistently test the
null hypothesis that a population density is $\rho-$concave using
any consistent shape constrained density estimator. This can be done
by estimating the population distribution subject to the shape constraint
and then using one of the classic tests for whether or not the empirical
distribution of the data is equal to this estimate; see for example
(Smirnov, 1948; Anderson and Darling, 1952). In these cases, choosing
a test with a statistic that is closely related to the fidelity criterion
used for estimation allows for a more straightforward interpretation
of the result. For example, if the test statistic is equal to the
fidelity criterion that the estimator optimizes, we would reject the
null if and only if we would also reject the null for every density
that satisfies the shape constraint. 

The following theorem provides the distribution of $W_{\gamma}(f_{unc},\mu)$
and a consistent test for the null hypothesis that $\mu^{\star}(x)$
satisfies the shape constraint based on this distribution. The method
also has the straightforward interpretation from the preceding paragraph,
so, if we denote the set of $\rho-$concave densities by $\mathcal{K}_{\rho},$
the null is rejected $\min_{f\in\mathcal{K}_{\rho}}W_{\gamma}(f,\mu)-W_{\gamma}(f_{unc},\mu)$
is statistically significant. Since $W_{\gamma}(f,\mu)$ is differentiable
in $f,$ this can be achieved without conditioning on the set of active
constraints. 

$\vphantom{c}$ 

\textbf{Theorem 5:}\textit{ Suppose the assumptions from Lemma 1 hold
and that $\sigma N\rightarrow\infty.$ Let $T$ be defined as $N(\sigma^{2}+\gamma/2)^{d/2}\left(W_{\gamma}(\mu^{\star},\mu)-W_{\gamma}(f_{unc},\mu)\right),$
$\psi$ as the optimal coupling between $\mu$ to $f_{unc},$ $\psi^{\star}$
as the optimal coupling between $\mu^{\star}$ and itself, $B^{\star}$
as $\gamma(D_{\mu^{\star}}-\psi^{\star}D_{\mathbf{1}\oslash\mu^{\star}}\psi^{\star T})^{+}/2,$
and $B$ as $\gamma(D_{f_{unc}}-\psi D_{\mathbf{1}\oslash\mu}\psi^{T})^{+}/2.$
Then, $T\overset{d}{\rightarrow}Z^{T}B^{\star}Z,$ where the iid elements
of $Z\in\mathbb{R}^{m}$ are distributed $Z_{i}\sqrt{N(\sigma^{2}+\gamma/2)^{d/2}}\sim N\left(\triangle\mu_{i}^{\star}/2,\mu_{i}^{\star}/\left(2\sqrt{\pi}\right)^{d}\right).$ }

\textit{Also, the hypothesis that $\mu^{\star}$ satisfies the shape
constraint can be consistently tested at a significance level of $\alpha$
by rejecting the null when $N(\sigma^{2}+\gamma/2)^{d/2}\left(W_{\gamma}(\hat{f},\mu)-\right.$
$\left.W_{\gamma}(f_{unc},\mu)\right)\geq c_{\alpha},$ where $c_{\alpha}$
satisfies $P(X^{T}BX\geq c_{\alpha})=\alpha$ and the iid elements
of $X\in\mathbb{R}^{m}$ are distributed $X_{i}\sqrt{N(\sigma^{2}+\gamma/2)^{d/2}}\sim N\left(0,\mu_{i}/\left(2\sqrt{\pi}\right)^{d}\right).$ }

\textit{Proof:}

The gradient and Hessian of $W_{\gamma}(f,\mu)$ with respect to $f$
are $\triangledown_{f}W_{\gamma}(f,\mu)=x_{\mu,f}$ and $\triangledown_{f,f}W_{\gamma}(f,\mu)=\triangledown_{f}x_{\mu,f}=\gamma(D_{f}-\psi_{\mu,f}D_{\mathbf{1}\oslash\mu}\psi_{\mu,f}^{T})^{+},$
which are derived in the proof of the next theorem. Thus, the second
order Taylor series expansion about $\mu^{\star}=f_{unc}$ is given
by,

$W_{\gamma}(\mu^{\star},\mu)=W_{\gamma}(f_{unc},\mu)+x_{\mu,f_{unc}}^{T}(\mu^{\star}-f_{unc})$
\begin{center}
$+(f_{unc}-\mu^{\star})^{T}B(f_{unc}-\mu^{\star})+O(\left\Vert f_{unc}(\mu)-\mu^{\star}\right\Vert ^{3}).$
\par\end{center}

Since $f_{unc}:=\text{arg min}_{f}W_{\gamma}(f,\mu)$ and $\triangledown_{f}W_{\gamma}(f,\mu)=x,$
we have $x=\mathbf{0}.$ Given the general discretized densities $\mu_{0},\mu_{1}\in\mathbb{R}^{m},$
the next theorem will also shows that the matrix \textit{$D_{\mu_{0}}-\psi D_{\mathbf{1}\oslash\mu_{1}}\psi^{T}$}
has one eigenvalue that is zero and $m-1$ eigenvalues that are strictly
positive. Note that the pseudo inverse is a continuous function when
its domain is restricted to the set of matrices with the same rank
(Stewart, 1969). Since $\mu$ and $f_{unc}$ converge in probability
to $\mu^{\star},$ the Slutsky theorem implies $B\overset{p}{\rightarrow}B^{\star}.$
Lemma 1 implies that $N(\sigma^{2}+\gamma/2)^{d/2}\left\Vert f_{unc}-\mu^{\star}\right\Vert ^{3}$
converges to zero in probability at a rate of $O_{p}\left(N^{-1/2}(\sigma^{2}+\gamma/2)^{-d/4}\right).$
Combining these results with the limiting distribution of $f_{unc}$
given in Lemma 1 implies that the Taylor series expansion given above
can be written as,
\begin{center}
$T:=N(\sigma^{2}+\gamma/2)^{d/2}\left(W_{\gamma}(\mu,\mu^{\star})-W_{\gamma}(f_{unc},\mu)\right)\overset{d}{\rightarrow}Z^{T}B^{\star}Z.$
\par\end{center}

\noindent Since $\mu$ is a consistent estimator for $\mu^{\star},$
we also have $X\overset{d}{\rightarrow}Z,$ so we also have $T\overset{d}{\rightarrow}X^{T}B^{\star}X.$ 

To show the test is consistent, suppose $\mu^{\star}(x)$ is not $\rho-$concave.
Recall that $W_{\gamma}(\hat{f},\mu)$ converges to the metric $W_{0}(\hat{f}(x),\mu(x))$
asymptotically (Benamou et al, 2015) and that Wasserstein distance
metrizes weak convergence of distributions (and convergence in the
first two moments). Continuity of the functions $\hat{f}(x),\mu(x)$
and $\mu^{\star}(x)$ implies that the distributions corresponding
to these densities converge weakly if and only if the densities themselves
converge to one another in probability. Since $\hat{f}$ is in the
feasible set and $\mu^{\star}$ is not, $\hat{f}(x)$ cannot converge
to $\mu^{\star}(x)$ in probability, so $W_{0}(\mu^{\star},\hat{f})$
does not converge to zero. Also, asymptotically we have $W_{0}(\hat{f},\mu)-W_{0}(\mu^{\star},\mu)>W_{0}(\mu^{\star},\hat{f})$
by the triangle inequality, so $N(\sigma^{2}+\gamma/2)^{d/2}\left(W_{\gamma}(\hat{f},\mu)-W_{\gamma}(f_{unc},\mu)\right)$
diverges under the alternative hypothesis. 
\begin{flushright}
$\square$
\par\end{flushright}

\section{The Optimization Algorithm}

In this section we will derive a trust region algorithm to find the
global minimum of (13) and (14). To do this, we will require the gradient
and the Hessian of $W_{\gamma}(\mu,g^{1/\rho}).$ The following Theorem
provides these values and shows that the optimization problem is convex
for cases in which $\rho\neq0.$ Appendix A contains these derivations
for the $\log-$concave case, which corresponds to the limit as $\rho\rightarrow0.$
For notational convenience, the Hessian given below corresponds to
the case in which the index $k$ is set equal to $m;$ although this
is not a requirement of the theorem. 

$\vphantom{c}$ 

\textbf{Theorem 6:}\textit{ The gradient of the $W_{\gamma}(\mu,g^{1/\rho})$
is}

\textit{
\begin{equation}
r:=\triangledown_{g}W_{\gamma}(\mu,g^{1/\rho})=D_{g^{1/\rho-1}/\rho}(x_{-k}-x_{k}),
\end{equation}
}

\noindent \textit{and the Hessian is }

\textit{
\begin{equation}
H:=\triangledown_{g}^{2}W_{\gamma}(\mu,g^{1/\rho})=ABA^{T}+C,
\end{equation}
}

\noindent \textit{where $A:=\left[\begin{array}{cc}
D_{g^{1/\rho-1}/\rho} & -g^{1/\rho-1}/\rho\end{array}\right],$ $B:=\gamma(D_{g^{1/\rho}}-\psi D_{\mathbf{1}\oslash\mu}\psi^{T})^{+},$
and $C:=\frac{1-\rho}{\rho^{2}}$ $\begin{array}{c}
D_{g^{1/\rho-2}}\end{array}D_{x_{-k}-x_{k}}.$ In addition, $W_{\gamma}(f,\mu)$ is strictly convex in $f$ when
$k$ is chosen to be $\text{arg min}_{i}\;x_{i}$ and $\gamma>0,$
and the optimization problem given in (13)-(14) is convex.}

\textit{Proof:} Since $W_{\gamma}(\mu,g^{1/\rho})$ is differentiable,
the envelope theorem implies that the gradient of the objective function
in (12) is equal to the gradient of the function given in (13), so
$\triangledown_{g}W_{\gamma}(\mu,g^{1/\rho})=D_{g^{1/\rho-1}/\rho}(x_{-k}-x_{k}).$ 

The derivative of (26) yields the sum of two matrices. Specifically,
\begin{center}
$H=\triangledown_{g}^{2}W_{\gamma}(\mu,f(g))$
\par\end{center}

\begin{equation}
=D_{g^{1/\rho-1}/\rho}\triangledown_{g}(x_{-k}(g)-x_{k}(g))+\left(\triangledown_{g}g^{1/\rho-1}/\rho\right)D_{x_{-k}-x_{k}}.
\end{equation}

\noindent Next we will begin by deriving the first term, which will
require several intermediate derivatives.

First, since, $f=(g^{1/\rho},m-\mathbf{1}\cdot g^{1/\rho}),$ we have 
\begin{center}
$\triangledown_{g}f(g)=\left[\begin{array}{c}
D_{g^{1/\rho-1}/\rho}\\
-g^{1/\rho-1}/\rho
\end{array}\right].$ 
\par\end{center}

\noindent Second, we will view $w$ as a function of $f$ in order
to find $\triangledown_{f}w(f).$ (9) and (10) can be combined to
yield the equality $f-D_{w}K\left(\mu\oslash(Kw)\right)=0,$ and implicit
differentiation of this equality implies,
\begin{center}
$\triangledown_{f}w(f)=(D_{f\oslash w}-\psi D_{\mathbf{1}\oslash\mu}\psi^{T}D_{\mathbf{1}\oslash w})^{+}.$
\par\end{center}

\noindent Third, the definition $w:=\exp\left(x/\gamma\right)$ implies
$x=\gamma\log(w),$ so 
\begin{center}
$\triangledown_{w}x(w)=\gamma D_{\mathbf{1}\oslash w}.$
\par\end{center}

\noindent Lastly, let $\tilde{x}(x):=x_{-k}-x_{k},$ so $\triangledown_{x}\tilde{x}(x)=\left[\begin{array}{cc}
I & \;-\mathbf{1}\end{array}\right]_{m-1\times1}.$ After combining all four derivatives, we have 
\begin{center}
$D_{g^{1/\rho-1}/\rho}\triangledown_{g}\tilde{x}(x(w(f(g))))=D_{g^{1/\rho-1}/\rho}\triangledown_{x}\tilde{x}(x)\triangledown_{w}x(w)\triangledown_{f}w(f)\triangledown_{g}f(g)$
\par\end{center}

\begin{center}
$=\gamma D_{g^{1/\rho-1}/\rho}\left[\begin{array}{cc}
I & \;-\mathbf{1}\end{array}\right]D_{\mathbf{1}\oslash w}(D_{f\oslash w}-\psi D_{\mathbf{1}\oslash\mu}\psi^{T}D_{\mathbf{1}\oslash w})^{+}\left[\begin{array}{c}
D_{g^{1/\rho-1}/\rho}\\
-g^{1/\rho-1}/\rho
\end{array}\right]$
\par\end{center}

\begin{center}
$=\gamma\left[\begin{array}{cc}
D_{g^{1/\rho-1}/\rho} & \;-g^{1/\rho-1}/\rho\end{array}\right](D_{f}-\psi D_{\mathbf{1}\oslash\mu}\psi^{T})^{+}\left[\begin{array}{c}
D_{g^{1/\rho-1}/\rho}\\
-\left(g^{1/\rho-1}\right)^{T}/\rho
\end{array}\right]$ 
\par\end{center}

\begin{center}
$=ABA^{T}.$
\par\end{center}

\noindent Since $\triangledown_{g}g^{1/\rho-1}/\rho=(1-\rho)/\rho^{2}D_{g^{1/\rho-2}},$
the second matrix in (28) is given by $C.$ 

Convexity requires that this Hessian is positive semidefinite. If
$k$ is chosen to be $\text{arg min}_{i}\;x_{i},$ then $x_{-k}-x_{k}\geq0.$
Since this is the case, $C$ is a diagonal matrix with nonnegative
diagonal elements, so $C$ is positive semidefinite. 

Next we will establish that $ABA^{T}$ is positive definite in several
steps. First, note that $ABA^{T}$ is symmetric if $B$ is symmetric.
Also, since $D_{f}$ and $\psi D_{\mathbf{1}\oslash\mu}\psi^{T}$
are symmetric $B^{+}$ is also symmetric. Since the Moore-Penrose
pseudo inverse preserves symmetry, $B$ is also symmetric. We will
proceed by establishing a few intermediary results on the components
of $ABA^{T}.$ 

Since $D_{\mathbf{1}\oslash\mu}$ is positive semidefinite, $\psi D_{\mathbf{1}\oslash\mu}\psi^{T}$
is as well. Since $\psi$ is a coupling of the densities $\mu$ and
$f,$ we have
\begin{center}
$D_{\mathbf{1}\oslash f}\psi D_{\mathbf{1}\oslash\mu}\psi^{T}\mathbf{1}$
\par\end{center}

\begin{center}
$=D_{\mathbf{1}\oslash f}\psi\mathbf{1}=\mathbf{1}.$
\par\end{center}

\noindent In other words, $\mathbf{1}$ is an eigenvector of $D_{\mathbf{1}\oslash f}\psi D_{\mathbf{1}\oslash\mu}\psi^{T}$
with a corresponding eigenvalue of 1. The Perron-Frobenius theorem
states that an $m\times m$ matrix with all positive elements and
columns that sum to one has a unique eigenvalue that is equal to one
and $m-1$ eigenvalues that are strictly less than one. Note that
each eigenvalue, say $\lambda,$ and corresponding eigenvector, say
$p,$ of $D_{\mathbf{1}\oslash f}\psi D_{\mathbf{1}\oslash\mu}\psi^{T}D_{\mathbf{1}\oslash f}$
satisfies,
\begin{center}
$\left(D_{\mathbf{1}\oslash f}\psi D_{\mathbf{1}\oslash\mu}\psi^{T}-\lambda I\right)p=0,$
\par\end{center}

\begin{center}
$\implies\left(I-D_{\mathbf{1}\oslash f}\psi D_{\mathbf{1}\oslash\mu}\psi^{T}-\tilde{\lambda}I\right)p=0$
\par\end{center}

\noindent so, if $p$ is an eigenvector of $D_{\mathbf{1}\oslash f}\psi D_{\mathbf{1}\oslash\mu}\psi^{T}D_{\mathbf{1}\oslash f},$
then $p$ is an eigenvector of $I-D_{\mathbf{1}\oslash f}\psi D_{\mathbf{1}\oslash\mu}\psi^{T}D_{\mathbf{1}\oslash f},$
with an eigenvalue corresponding to $\tilde{\lambda}:=1-\lambda.$
This implies that $I-D_{\mathbf{1}\oslash f}\psi D_{\mathbf{1}\oslash\mu}\psi^{T}$
is a positive semidefinite matrix with rank $m-1.$ Since multiplication
by a positive definite matrix and applying the pseudoinverse preserve
both the rank and the signs of the eigenvalues, $B=\gamma\left(D_{f}\left(I-D_{\mathbf{1}\oslash f}\psi D_{\mathbf{1}\oslash\mu}\psi^{T}\right)\right)^{+}$
is also a positive semidefinite matrix with rank $m-1.$ 

Observation 7.1.8 in Horn and Johnson (1990) implies that the nullspace
of $ABA^{T}$ is the same as the nullspace of $BA^{T}.$ Since the
eigenvector of $B$ that corresponds to the eigenvalue of zero is
$\mathbf{1},$ $ABA^{T}$ is positive definite if there does not exist
$v\in\mathbb{R}^{m-1}$ such that $A^{T}v=\mathbf{1}_{m}.$ This system
of equations is equivalent to $g_{i}^{1/\rho-1}v_{i}=\rho$ for all
$i\in\{1,...,m-1\}$ and $\sum_{i}g_{i}^{1/\rho-1}v_{i}=-\rho,$ which
does not have a solution, so $ABA^{T}$ is positive definite. 

This, along with the fact that the constraints in (14) are equivalent
to constraining $sgn(\rho)g$ to be in the set of concave functions,
which is a convex cone, implies that the optimization problem is convex. 
\begin{flushright}
$\square$
\par\end{flushright}

\textbf{Remark:} Choosing $k$ to be \textit{$\text{arg min}_{i}\;x_{i}$}
is a sufficient, but not necessary, condition to guarantee convexity.
Choosing $k$ to correspond to the element on the boundary of the
mesh over $\mathcal{A},$ with the lowest corresponding value of $w_{i},$
ensures that the density estimate satisfies the shape constraints
everywhere on the interior of its domain and rarely results in the
objective function being nonconvex along the convergence path. 

As described below, we initialize our algorithm at a very good approximation
of $\hat{f}.$ Note that this method does not require the specification
of $k.$ When the mesh is enlarged beyond the convex hull of the data,
so that $\mu_{i}$ and $\hat{f}_{i}$ are lowest when $\mathbf{a}_{i}$
corresponds to a point on the boundary of the domain, $w$ and $v$
generally obtain their minimum value on the boundary points. Thus,
initializing the algorithm at a density that is near $\hat{f}$ almost
always results in $w$ falling to a sufficiently low value at points
corresponding to the boundary of $\mathcal{A}$ to ensure strict positive
definiteness of $H$ along the entire convergence path.\footnote{It may be possible to modify the algorithm we use to compute the approximation,
as described in Appendix B, to provide a convergence guarantee. This
would allow for the formulation of the estimator using $g\in\mathbb{R}^{m}$
rather than $g\in\mathbb{R}^{m-1}.$ This is an ongoing area of research. }
\begin{flushright}
$\square$
\par\end{flushright}

Having already derived the gradient and Hessian of $W_{\gamma}(\mu,g^{1/\rho}),$
it is straightforward to create a trust region algorithm. The algorithm
takes an initial density estimate, $f^{(0)},$ as input and in iteration
$i$ the algorithm solves 
\begin{center}
$\;\;\;\;\Delta\leftarrow\text{arg min}_{d}\;\;\left\{ d^{T}Hd/2+d^{T}r\;\mid\;\left(d+g_{-k}^{(i-1)}\right)sgn(\rho)\in\mathcal{K},\;\left\Vert d\right\Vert \leq c\right\} .$
\par\end{center}

\noindent If the value of the objective function evaluated at $g_{-k}^{(i-1)}+\Delta$
results in an improvement over its value at $g_{-k}^{(i-1)},$ then
$g_{-k}^{(i)}$ is defined to be $g_{-k}^{(i-1)}+\Delta.$ If the
improvement was significant, then the radius of the trust region,
$c,$ is increased, and otherwise it is decreased and the value of
$g_{-k}^{(i)}$ is defined to be $g_{-k}^{(i-1)}.$ This is described
in Algorithm 2. 
\begin{algorithm}[H]
\textbf{function}\texttt{ Trust Region}($\mu,\rho,f^{(0)},K$)

$(W_{\gamma}^{(0)},H^{(0)},r^{(0)},k^{(0)})\leftarrow$\texttt{Wasserstein
Distance($\mu,\rho,K,f^{(0)}$)}

$g_{-k}^{(0)}\leftarrow f_{-k}^{(0)^{\rho}}$

$\tilde{W}_{\gamma}^{(0)}\leftarrow W_{\gamma}^{(0)}$

$\eta\leftarrow1,$ $c\leftarrow1$

\textbf{for $i=1,2,...$: }

$\;\;\;\;$$(W_{\gamma}^{(i)},H^{(i)},r^{(i)},k^{(i)})\leftarrow$\texttt{Wasserstein
Distance($\mu,\rho,K,f^{(i-1)}$)}

$\;\;\;\;$\textbf{if $W_{\gamma}^{(i)}>W_{\gamma}^{(i-1)}$:} $W_{\gamma}^{(i)}\leftarrow W_{\gamma}^{(i-1)},\;H^{(i)}\leftarrow H^{(i-1)},$

$\;\;\;\;\;\;\;\;$ $r^{(i)}\leftarrow r^{(i-1)},\;k^{(i)}\leftarrow k^{(i-1)},\;g\leftarrow f^{(i-2)^{\rho}}$

$\;\;\;\;$\textbf{if $(W_{\gamma}^{(i)}-W_{\gamma}^{(i-1)})/\tilde{W}_{\gamma}^{(i-1)}<0.25$:}
$c\leftarrow c/4+\eta/8$

$\;\;\;\;$\textbf{else: }$c\leftarrow3c/2$

$\;\;\;\;$$\Delta\leftarrow\text{arg min}_{d}\;\;d^{T}Hd/2+d^{T}r\;\;\text{s.t.}\;\;d+g\in\mathcal{K},\;\left\Vert d\right\Vert \leq c.$

$\;\;\;\;$$\tilde{W}_{\gamma}^{(i)}\leftarrow\Delta^{T}H\Delta/2+\Delta^{T}r$

$\;\;\;\;$$\eta\leftarrow\left\Vert \Delta\right\Vert $

$\;\;\;\;$$g\leftarrow g+\Delta$

$\;\;\;\;$$g_{k}\leftarrow(m-\underset{i}{\sum}g_{i}^{1/\rho})^{\rho}$ 

$\;\;\;\;$$f^{(i)}\leftarrow g^{1/\rho}$

\textbf{return $f$}

\caption{The parameter values for this method were set using the recommendations
of Fan and Yuan (2001). \texttt{Wasserstein Distance$(\cdot)$} uses
Algorithm 1 to find $w,$ and then computes $W_{\gamma}(\mu,f),$
the gradient, Hessian, and the index $k$ (as defined in Theorem 6).}
\end{algorithm}

Figures 1 and 2 illustrate two examples of the output of Algorithm
2. Figure 1 provides density estimates of the rotational velocity
of stars that are constrained to be $-2-$concave and $-1/2-$concave,
respectively. Figure 2 provides a plot of a two dimensional density;
to illustrate the tail behavior of the density more clearly, the logarithm
of the density is shown. This estimator uses a dataset containing
the height and left middle finger length of 3,000 British criminals
that was analyzed by Macdonell (1902) and Student (1908).
\begin{figure}[H]
\begin{centering}
\subfloat[]{\begin{centering}
\includegraphics[scale=0.6]{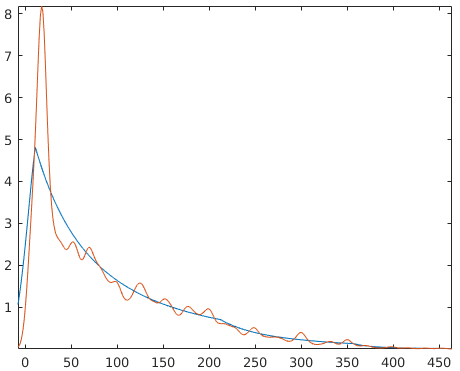}
\par\end{centering}
}\subfloat[]{\begin{centering}
\includegraphics[scale=0.6]{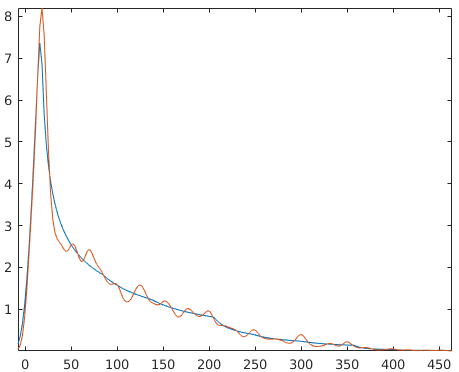}
\par\end{centering}
}
\par\end{centering}
\caption{The red density in each plot is a kernel density estimator of the
rotational velocity of stars from from Hoffleit and Warren (1991).
The blue density is the algorithm's output. $\rho$ was set equal
to $-0.5$ in (a) and $-2$ in (b). $\gamma$ was set using using
the first method described above, so the constraints are binding almost
everywhere. }
\end{figure}

The data used to generate both figures are also used by Koenker and
Mizera (2010) to compare $\log-$concave density estimates with $-1/2-$concave
density estimates. In the case of the dataset for the rotational velocity
of stars, they show that the former provides a monotonic density in
the region in which the speed of rotation is strictly positive, while
the latter density has a peak near the mode of the kernel density
estimator shown in Figure 1. This peak is also present in both of
the shape-constrained densities shown in Figure 1. 

For the dataset used in Figure 2, Koenker and Mizera (2010) show that
the logarithm of the maximum likelihood density estimator subject
to a $\log-$concavity constraint is below $-24$ near the observation
at the very top of Figure 2, so observations this far from the rest
of the data would be fairly unlikely to occur if the density was in
fact $\log-$concave. The logarithm of the $-1/2-$concave density
given below is approximately $-7.2$ near this observation.
\begin{figure}[H]
\begin{centering}
\includegraphics[scale=0.61]{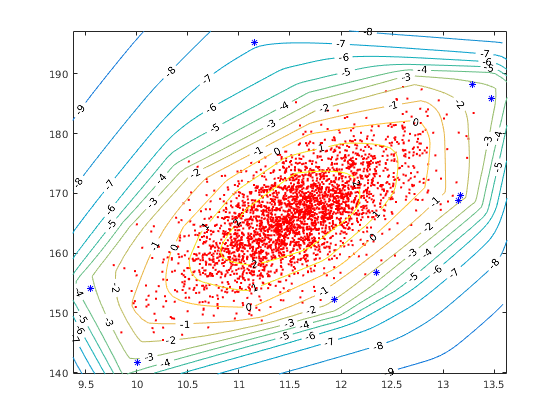}
\par\end{centering}
\caption{The data points (shown in red) consists of the height and finger length
of 3,000 criminals from Macdonell (1902). The points on the convex
hull of the data are illustrated with blue asterisks. The contour
plot depicts the logarithm of the $-1/2-$concave density estimate
to illustrate the tail behavior of the density. }
\end{figure}

To compute these estimates as well as the estimates given in the application
section, each iteration of Algorithm 2 used MOSEK, a highly optimized
quadratic program solver, to find $\Delta$ efficiently; however,
this step is the limiting variable in terms of the time complexity
of the algorithm. The computational efficiency of the algorithm can
be improved by initializing $f^{(0)}$ at a good approximation of
the final density estimate. Also, using the best available approximation
of $\hat{f}$ to define $f^{(0)}$ ensures that $w$ is as close as
possible to $\mathbf{1},$ for the reasons given in Remark 1 of Theorem
2. This results in an increase in the numerical accuracy of the gradient
and Hessian in Algorithm 2, and thus increased stability of the algorithm.
The appendix provides a method for finding a particularly accurate
and computationally efficient approximation. In many cases initializing
Algorithm 2 at this input density results in convergence within two
to four iterations, bringing the total computation time of our implementation
down to approximately 10 seconds in our MATLAB implementation and
35 seconds in our R implementation when $m=200.$\footnote{We have not finished optimizing these implementations. We believe
there is an opportunity for an improvement in this computation time
by at least an order of magnitude, particularly in the version written
in R. }

The next section provides results on a generalization of the optimization
problem given in (13)-(14). One of these results provides sufficient
conditions for convexity of this more general optimization problem.
It is worth noting that a similar sequential quadratic programming
algorithm can also be used to find the global optimum in this more
general setting when the derivatives corresponding to $H$ and $r$
exist. Alternatively, one could use gradient descent to solve this
more general optimization problem, which would only require the user
to provide the first derivative of the transformation of $g.$

\section{Shape-Constraints More Generally}

Although $\log-$concavity, and $\rho-$concavity more generally,
are the most commonly studied shape constraints, the results provided
in the previous sections are also applicable to a large class of new
shape constraints. Specifically, this section considers solutions
to the general optimization problem given by,

\begin{equation}
\underset{g}{\min}\;W_{\gamma}(\mu,\alpha(g))
\end{equation}

\begin{equation}
\text{subject to: }\cap_{i}\beta_{i}(g)\leq0,
\end{equation}

\noindent where $\alpha_{i}:\mathbb{R}^{m-1}\rightarrow\mathbb{R}_{+}$
and $\beta_{i}:\mathbb{R}^{m-1}\rightarrow\mathbb{R}.$ Analogous
to the estimator described in the preceding sections, we will denote
the minimizer of (29)-(30) as $\tilde{g}$ and the generalized density
estimator as $\tilde{f}:=f(g),$ where $f(g):=[\alpha(g)^{T},m-\mathbf{1}^{T}\alpha(g)]^{T}.$
The following theorem provides sufficient conditions for the results
provided in Theorems 2, 4, and 5 to hold in this more general setting.
The statements of these theorems were purposefully ambiguous in regards
to the constraint, so we simply provide additional requirements on
the functions $\alpha(\cdot)$ and $\beta(\cdot)$ for these generalizations.
We also provide sufficient conditions for Theorem 6, but this requires
restating the result in its entirety using the new notation. Lastly,
point (4) of the theorem provides a new result for strict convexity
of $W_{\gamma}(\mu,\alpha(g))$ in the neighborhood of its global
minimum, without requiring the assumption that $\alpha(\cdot)$ is
convex or concave. 

$\vphantom{c}$ 

\textbf{Theorem 7:}\textit{ Let $i(x)$ be defined as $\text{arg min}_{i}\left\Vert x-\mathbf{a}_{i}\right\Vert ,$}
\textit{$\Theta$ as the set of proper and uniformly continuous density
functions, $\Omega_{N}$ as $\{g\mid\cap_{i}\beta_{i}(g)\leq0\land\alpha(g)\in\mathbb{R}^{m}\land\sum_{i}\alpha_{i}(g)\leq m\},$
the set $\Lambda$ so that $f(x)\in\Lambda$ if and only if there
exists a sequence $\{g^{(N)}\}_{N}^{\infty}$ such that $g^{(N)}\in\Omega_{N}$
for all $N$ sufficiently large and $f(x)=\lim_{N\rightarrow\infty}[\alpha(g^{(N)})^{T},m-\mathbf{1}^{T}\alpha(g^{(N)})]_{i(x)}.$ }

\textit{If $\mu^{\star}(x)\in\Theta,$ both of the sets $\Theta\cap\Lambda$
and $\Omega_{N}$ are nonempty, the function $\alpha(\cdot)$ is invertible,
then the following additional conditions are sufficient for the applicability
of Theorems 2, 4, 5, and 6 to $\tilde{f}.$ }

\textit{(1) If the codomain of the function }$g\mapsto[\alpha(g)^{T},m-\mathbf{1}^{T}\alpha(g)]$\textit{
contains $f_{unc}$ then Theorems 2 and 5 hold for $\tilde{f}.$ }

\textit{(2) If each of the functions $\beta_{i}(g),$ as well as $\alpha(g),$
are differentiable in the neighborhood of $\tilde{g}$ asymptotically,
$\tilde{g}$ converges to a point on the interior of the domains of
these functions, and $\triangledown_{g}\beta_{i}(g)\mid_{g=\tilde{g}}\neq0$
for each $i$ and $\triangledown_{g}\alpha(g)\mid_{g=\tilde{g}}\neq0,$
then Theorem 4 holds for $\tilde{f}.$ }

\textit{(3) Suppose $\{g\mid\beta_{i}(g)\leq0\}$ is convex for all
$i,$ and $\alpha_{j}(g)$ is convex (respectively, concave) for all
$j.$ If $k:=\text{arg min}_{i}\;x_{i}$ ($k:=\text{arg max}_{i}\;x_{i}$),
then (29)-(30) is a convex optimization problem. Also, the gradient
and Hessian of $W_{\gamma}(\mu,\alpha(g))$ exist almost everywhere,
and at these points they are given by}
\begin{center}
$\triangledown_{g}W_{\gamma}(\mu,\alpha(g))=(x_{-k}-x_{k})\triangledown_{g}\alpha(g),$
\par\end{center}

\begin{center}
$\triangledown_{g}^{2}W_{\gamma}(\mu,\alpha(g))=ABA^{T}+C,$
\par\end{center}

\noindent \textit{where} \textit{$A:=\left[\begin{array}{cc}
\triangledown_{g}\alpha(g) & \;-\triangledown_{g}\alpha(g)\mathbf{1}\end{array}\right],$ $B:=\gamma(D_{\alpha(g)}-\psi D_{\mathbf{1}\oslash\mu}\psi^{T})^{+},$
and $C:=\sum_{j}(x_{j}-x_{k})\triangledown_{g}^{2}\alpha_{j}(g).$ }

\textit{(4) Let $d:\mathbb{R}^{m-1}\times\mathbb{R}^{m-1}\rightarrow\mathbb{R}_{+}^{1}$
denote an arbitrary distance measure. If $\alpha(g)\in C^{2}$ and
$\triangledown_{g}\alpha(g)$ has full rank, then there exists $\delta>0$
such that $W_{\gamma}(\mu,\alpha(g))$ is convex in $g$ when }$d([\alpha(g)^{T},m-\mathbf{1}^{T}\alpha(g)],f_{unc,-k})\leq\delta.$ 

\textit{Proof: (1):} The proof of theorems 2 and 5 do not use the
$\rho-$concavity constraint, so they still hold as long as there
exists $g_{unc}$ such that $f_{unc}=f(g_{unc}).$ Since $f_{unc}$
is the global minimum of $f\mapsto W_{\gamma}(f,\mu),$ $\tilde{f}$
will be equal to $f_{unc}$ whenever $g_{unc}$ is feasible. 

\textit{(2):} The proof given for Theorem 4 uses the first order delta
method, which requires the conditions given in the theorem. These
conditions are also sufficient for the application of the continuous
mapping theorem, which was used to show convergence in probability
of the estimator. 

\textit{(3): }Since each \textit{$\alpha_{j}(g)$ }is convex,\textit{
}Aleksandrov's theorem implies that $\triangledown_{g}\alpha_{j}(g)$
and $\triangledown_{g}^{2}\alpha_{j}(g)$ exist almost everywhere.
We will begin by deriving the gradient and Hessian at these points.
Let $\omega(x,y,g)$ be defined as 
\begin{center}
$x_{-k}^{T}\alpha(g_{-k})+x_{k}\left(m-\mathbf{1}_{m-1}^{T}\alpha(g_{-k})\right)+y^{T}\mu-\gamma\sum_{i,j}\exp\left((x_{i}+y_{j}-M_{ij})/\gamma\right),$
\par\end{center}

\noindent so that we can write $W_{\gamma}(\mu,\alpha(g))$ as 
\begin{center}
$\max_{x,y}\omega(x,y,g).$
\par\end{center}

\noindent Since $\omega(x,y;g)$ is differentiable in all of its arguments,
the envelope theorem implies 
\begin{center}
$\triangledown_{g}W_{\gamma}(\mu,\alpha(g))=\triangledown_{g}\omega(x,y,\alpha(g))=(x_{-k}-x_{k})\triangledown_{g}\alpha(g).$
\par\end{center}

\noindent By the same logic used in the proof of Theorem 4, we can
write the Hessian as,
\begin{center}
$\triangledown_{g}^{2}W_{\gamma}(\mu,\alpha(g))=\left(\triangledown_{g}f(g)\right)\left(\triangledown_{f}x(w(f))\right)\left(\triangledown_{g}f(g)\right)^{T}+\sum_{l\neq k}(x_{l}-x_{k})\triangledown_{g}^{2}\alpha_{l}(g)$
\par\end{center}

\begin{center}
$=ABA^{T}+C.$
\par\end{center}

$ABA^{T}$ is positive definite by the same argument used in Theorem
6, which is also expanded on further in the proof of statement (2).
When each $\alpha_{j}(g)$ is convex and $\triangledown_{g}^{2}\alpha_{j}(g)$
exists, then $\triangledown_{g}^{2}\alpha_{j}(g)$ is a positive semidefinite
matrix. In this case, $k:=\text{arg min}_{i}\;x_{i},$ so each element
of $(x_{-k}-x_{k}\mathbf{1})$ is nonnegative. This implies each term
in the sum defining $C$ is a positive semidefinite matrix, so $C$
is positive semidefinite. Also, when each $\alpha_{j}(g)$ is concave
and $\triangledown_{g}^{2}\alpha_{j}(g)$ exists, then $\triangledown_{g}^{2}\alpha_{j}(g)$
is negative semidefinite. Choosing $k:=\text{arg max}_{i}\;x_{i}$
implies $(x_{j}-x_{k})\triangledown_{g}^{2}\alpha_{j}(g)$ is a positive
semidefinite matrix, so $C$ is also positive semidefinite in this
case. 

Since \textit{$\{g\mid\beta_{i}(g)\leq0\}$} is convex for all \textit{$i,$}
the intersection of these sets is also convex. This, combined with
the strict convexity of $W_{\gamma}(\mu,\alpha(g))$ in $g,$ implies
convexity of the optimization problem given in (29)-(30). 

\textit{(4): }The argument given in Remark 1 following Theorem 2 implies
that each of the elements of $x$ are the same when $f(g)=f_{unc}.$
When this occurs, we have $x_{-k}-x_{k}=0,$ so $C=\mathbf{0}_{m-1\times m-1}$
when $f(g)=f_{unc}.$ 

The argument in the second to last paragraph of Theorem 6 implies
that $ABA^{T}$ is positive definite when $\triangledown_{g}\alpha(g)$
has full rank and there does not exist $v\in\mathbb{R}^{m-1}$ such
that $A^{T}v=\mathbf{1}_{m}.$ This system of equations requires $\left(\triangledown_{g}\alpha(g)\right)v=\mathbf{1}_{m-1}$
and $-\mathbf{1}_{m-1}^{T}\triangledown_{g}\alpha(g)v=1.$ However,
if $v$ satisfies $\left(\triangledown_{g}\alpha(g)\right)v=\mathbf{1}_{m-1},$
then $-\mathbf{1}_{m-1}^{T}\triangledown_{g}\alpha(g)v=1-m,$ so there
is not a solution to this system of equations. This, combined with
the fact that $\triangledown_{g}\alpha(g)$ has full rank, implies
$ABA^{T}$ is positive definite. 

Since $\alpha(g)\in C^{2},$ the eigenvalues of the $\triangledown_{g}^{2}W_{\gamma}(\mu,\alpha(g))$
are continuous in $g.$ Since we have shown that these eigenvalues
are strictly positive when $f(g)=f_{unc},$ continuity implies that
there exists $\delta>0$ such that all eigenvalues are nonnegative
when $d(f(g),f_{unc})\leq\delta.$ 
\begin{flushright}
$\square$
\par\end{flushright}

\subparagraph*{Remark 1:}

Some of the assumptions given above were made to simplify the exposition
rather than necessity. For example, we can replace assumptions regarding
differentiability and rank for all $g$ with similar assumptions in
the neighborhood of $\tilde{g}.$ The assumption regarding the existence
of the inverse of $\alpha(\cdot)$ is worth mentioning in particular.
Cases in which either $\alpha(\cdot)$ or $\alpha^{-1}(\cdot)$ cannot
be expressed in a closed form appear to be fairly common, but closed
form solutions are not a requirement of the theorem since they can
be replaced with their numerical counterparts. The application provided
in the next section is one example of this case. 
\begin{flushright}
$\square$
\par\end{flushright}

Mechanism design appears to be a particularly fruitful source for
applications of this generalized shape constrained density estimator.
For example, consider a private values auction model in which bidders
have valuations that are drawn from the density $f(x).$ Myerson (1981)
defines the virtual valuations function as $J_{f}(x):=x-(1-F(x))/f(x),$
and shows that if $J_{f}(x)$ is monotone increasing, then an auction
that awards the item to the highest bidder is optimal in the sense
of maximizing expected revenue. It is common in mechanism design to
make the stronger assumption that the hazard function, defined by
$f(x)/(1-F(x)),$ is increasing or the even stronger assumption that
$f(x)$ is $\log-$concave. McAfee and McMillan (1987) show that monotonicity
of $x-(1-F(x))/f(x)$ is equivalent to convexity of the function $g(x)=1/(1-F(x)),$
which is used in the the next section to formulate a density estimate
subject to the constraint that $f(x)$ satisfies Myerson's (1981)
regularity condition.\footnote{This example also demonstrates that $\alpha(\cdot)$ and $\beta(\cdot)$
do not need to be unique, since we could constrain the discretized
counterparts of $1/(1-F(x))$ to be convex, $\triangledown_{x}1/(1-F(x))$
to be monotonic, or $\triangledown_{x,x}1/(1-F(x))$ to be positive. }

In addition, Myerson and Satterthwaite (1983) show that bilateral
bargaining between a buyer and a seller will only result in trade
when the virtual valuation of the buyer and the virtual cost of the
seller, defined by $x+F(x)/f(x),$ are both increasing functions.
Note that we can define $H(x):=1-F(x)$ and $h(x):=H'(x)=-f(x)$ to
show that this last condition is equivalent to monotonicity of $x-(1-H(x))/h(x).$
A reformulation of McAfee and McMillan's (1987) condition for monotonicity
of $J_{f}(x)$ shows that this is equivalent to convexity of $g(x):=1/(1-H(x)).$
This allows for the formulation of this shape constraint in an analogous
manner as the method used to formulate the shape constraint in the
next section. 

It would also be interesting to explore constraining a density to
have an increasing hazard function. Wellner and Laha (2017) show that
this is equivalent to constraining $g(x)=-\log(1-F(x))$ to be convex.
In all three of the examples given above, guaranteeing that $C$ is
positive definite requires the density estimate to satisfy a set of
inequalities that do not appear to have an obvious interpretation.
Regardless, statement (2) in Theorem 5 implies that $W_{\gamma}(\mu,\alpha(g))$
is still convex as long as $f(g)$ is sufficiently close to $f_{unc}.$
Initializing the density near this unconstrained minimizer and then
checking for convexity in each iteration often results in local convexity
of $W_{\gamma}(\mu,\alpha(g))$ along the path of convergence.

Since the eigenvalues of the positive definite matrix $ABA^{T}$ are
increasing in $\gamma,$ $\sqrt{\gamma/2+\sigma^{2}}$ can also be
increased to ensure that $W_{\gamma}(\mu,\alpha(g))$ is convex over
a larger set. This has the added benefit of increasing the dispersion
of $f_{unc},$ which results in $f_{unc}$ moving closer to the feasible
set in the case of most shape constraints, including all the examples
discussed so far. In some cases ensuring convexity by increasing $\sqrt{\gamma/2+\sigma^{2}}$
may result in the density being too disperse. If this occurs, it would
be best to compare the resulting density estimate with an estimate
subject to a stronger constraint, that allows for the formulation
of a convex optimization problem, and check which density estimate
fits the data more closely. For example, Ewerhart (2013) shows that
a sufficient condition for a density to satisfy Myerson's (1981) regularity
condition is $\rho-$concavity for $\rho>-1/2,$ and a $\log-$concavity
constraint can be used to ensure that the hazard function is monotonic. 

Many other examples of constraints that are commonly imposed on densities
in economics are given by Ewerhart (2013). Even though the examples
cited in this paper all constrain $g(x)$ to be concave or convex,
this is by no means a requirement of Theorem 7. For example, one could
define a density estimate of the form given by (29)-(30) to estimate
densities subject to any of the examples of shape constraints that
are given by Ewerhart (2013). 

\section{Myerson's (1981) Regularity Condition}

The California Department of Transportation (Caltrans) uses first
price auctions to allocate construction contracts. In this section
we use data on the bids submitted to Caltrans in 1999 and 2000 to
explore whether or not this choice of auction format minimizes the
costs to the state of California. As discussed in the previous section,
if $f(x)$ is the density of private valuations for the bidders, with
a distribution function denoted by $F(x),$ and if the bidders are
risk neutral, Myerson (1981) shows that auctions that award the item
to the person with the highest bid are optimal when the virtual valuations
of the bidders is monotonically increasing. 

To examine whether this condition is plausible, we need to estimate
the valuations (or, in this case, marginal costs) of the construction
firms. Guerre, Perrigne, and Vuong (2000) used the fact that the best
response function of bidders in a first price sealed bid auction is
an increasing function of the bidders's valuations to show that the
valuation of bidder $i$ can be estimated by 

\begin{equation}
b_{i}+\frac{\hat{F}_{b}(b_{i})}{(l-1)\hat{f}_{b}(b_{i})},
\end{equation}

\noindent where $l$ is the numbers of bidders participating in the
auction, $b_{i}$ is $i$'s bid, $\hat{f}_{b}(\cdot)$ is a consistent
estimate of the density of bids, and $\hat{F}_{b}(\cdot)$ is its
corresponding distribution function. To control for the size of each
project, we normalize each bid by Caltrans's engineers's estimates
of the cost of each project before estimating $\hat{f}_{b}(\cdot)$
and $\hat{F}_{b}(\cdot)$ for each auction size. 

Bajari, Houghton and Tadelis (2006) use the same dataset to estimate
the costs of each firm. We follow a similar estimation strategy but
make some modifications because our focus is on the costs to the state
of California. Specifically, we did not subtract transportation costs
from the cost estimates or treat bids from small firms differently
than larger firms. Each bid consists of a unit cost bid on each item
that the contract requires, and the total bid is equal to the dot
product of the number of items required and the unit bid of each item.
If small modifications are made to the contract after it is awarded,
the final payment to the firm is equal to the dot product of the modified
quantities and the unit costs in the original bid. Bajari, Houghton
and Tadelis (2006) found evidence that firms are able to make accurate
forecasts of these final quantities, so we follow their recommendation
and replace the first term in (26) with the final amount that is paid
to the firm (normalized by the Caltran's engineers' estimate of the
project cost). We also exclude all auctions in which these modifications
resulted in a change in the payment received by the firm by more than
3\%. After excluding these auctions we were left with 1,393 bids.
Lastly, Hickman and Hubbard (2015) showed that the accuracy of the
valuations estimates can be improved by applying a boundary correction
to $\hat{f}_{b}(\cdot),$ which we also employed in our estimation
procedure.

After we estimated the valuations for each firm, we estimated $f_{unc}$
by setting $\sqrt{\gamma/2+\sigma^{2}}$ using Scott's (1992) rule
of thumb; however, the resulting virtual valuations function was not
monotonic. This could be an innocuous idiosyncrasy of the data or
it could be evidence that Caltran's choice of auction format is suboptimal. 

To investigate which possibility is more plausible, we find the proposed
density estimate subject to Myerson's (1981) regularity condition.
To define $\alpha(\cdot)$ we solved for $F(x)$ in the equation introduced
in the previous section, $g(x)=1/(1-F(x)).$ This derivative is $f(x)=mg'(x)/g(x)^{2},$
and after discretizing, we defined $\alpha_{j}(g)$ as $(g_{j}-g_{j+1})/g_{j}^{2}.$
The convexity of the objective function was maintained along the entire
path of convergence, without requiring that we increase $\sqrt{\gamma/2+\sigma^{2}}$
above the recommendation given in the third section. The input density
and the output of the algorithm are shown in Figure 3. 

We also performed the test described in Theorem 5. We failed to reject
the null hypothesis that the objective function, evaluated at $\hat{f},$
was equal to the objective function evaluated at its unconstrained
counterpart, $f_{unc},$ with a $p-$value of 0.93. This is similar
to the result of the Kolmogorov-Smirnov (1948) test and the Anderson-Darling
(1952) test for the null hypothesis that the sample was drawn from
the distribution function of $\hat{f};$ these tests also failed to
reject the null with $p-$values of 0.98 and 0.94, respectively. In
this case the constraints are inactive at all but 24 points in the
right tail in a mesh of 300 points.\footnote{Myerson's (1981) regularity condition can also be expressed as $f(x)^{2}+f'(x)(1-F(x))\geq0,$
so it is always satisfied when the density is increasing. In this
case, $f_{unc}$ decreases rapidly to the right of the mode, as shown
in Figure 3, so it is not in the set of feasible densities. }

Since the density already appears parsimonious, there is little reason
to decrease $\sqrt{\gamma/2+\sigma^{2}};$ however, in the interest
of comparing these three tests further, we also estimated the density
using Scott's (1992) rule of thumb multiplied by 1/2 rather than 2/3.
In this case the $p-$value of our test decreased to 0.32, while the
$p-$values of other two tests both increased to 0.99. This divergence
in $p-$values underlines the difference between these two approaches.
Specifically, as we decrease the smoothing, the distribution function
converges to the empirical distribution function over the vast majority
of the domain, so tests based on comparisons between a distribution
and its empirical counterpart are less likely to reject. In contrast,
our statistic measures the discrepancy between the global unconstrained
minimum of $f\mapsto W_{\gamma}(\mu,f)$ and the set of feasible densities.
The test is most reliable when $f_{unc}$ is a reasonable estimate
of $\mu^{\star},$ so we do not recommend setting $\sqrt{\gamma/2+\sigma^{2}}$
to a value that under-smooths $f_{unc}$ in this way. 

\begin{figure}[H]
\begin{centering}
\subfloat[]{\begin{centering}
\includegraphics[width=7.3cm,height=7.3cm]{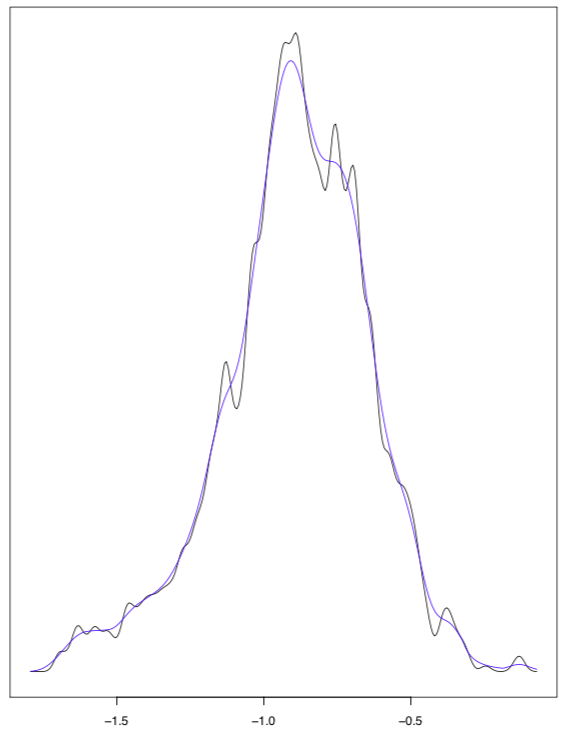}
\par\end{centering}
}\subfloat[]{\begin{centering}
\includegraphics[width=7.3cm,height=7.3cm]{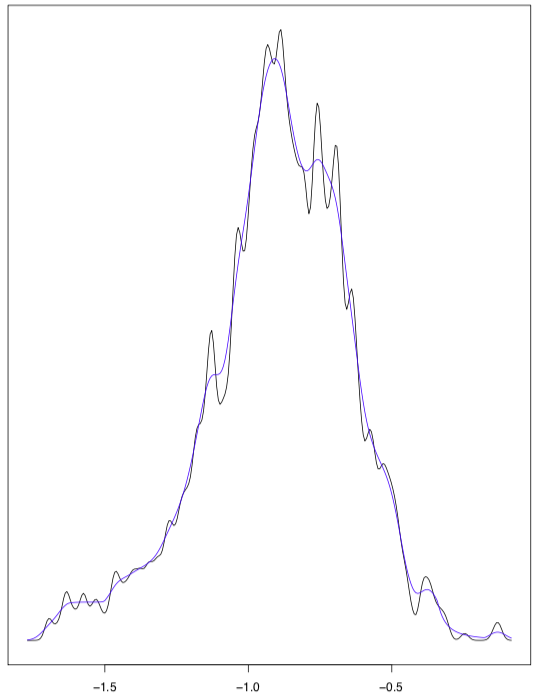}
\par\end{centering}
}
\par\end{centering}
\caption{The input density and the output density, subject to Myerson's (1981)
regularity condition, are shown in black and blue, respectively. (a)
shows the estimate when $\sqrt{\gamma/2+\sigma^{2}}$ is set using
Scott's (1992) rule of thumb multiplied by 2/3, while in (b) a factor
of 1/2 is used instead. The absolute value of the valuations can be
viewed as the cost per dollar of the Caltrans\textquoteright s engineer\textquoteright s
estimates.}
\end{figure}

\section{Conclusion }

This paper proposes a density estimator that is defined as the density
that minimizes a regularized Wasserstein distance from the input kernel
density estimator subject to $\rho-$concavity constraints. This framework
provides the advantages of convexity and consistency, and it allows
for a generalization that is capable of estimating densities subject
to a large class of alternative shape constraints. In addition, it
allows for a test of the impact of the shape constraints on the fidelity
criterion. 

The framework presented here can also be extended to allow $\gamma$
and $\sigma$ to take different values at each column of the matrix
$K,$ which would be appealing in two situations. When one would like
$f^{\star}$ to be as close as possible to $\mu,$ $\gamma$ and $\sigma$
can be decreased below what would have otherwise been possible in
regions where the shape-constrained density estimator is closer to
$\mu,$ without interfering with convergence of Algorithm 1. Secondly,
this would allow for the development of methods that set $\sqrt{\gamma/2+\sigma^{2}}$
using an adaptive approach that is similar to the one described by
Sheather and Jones (1991) for kernel density estimators. 

Another promising area for future research that has not already been
mentioned would be to extend this framework to allow for the estimation
of a regression and the density of residuals simultaneously. Dümbgen,
Samworth, and Schuhmacher (2011) showed that this does not result
in a convex optimization problem in the maximum likelihood setting,
so verifying convexity of the objective function in this case is an
active area of research. Note that extending the framework presented
here to estimating the mode of a data generating process conditional
on covariates, or a modal regression, is straightforward. For example,
this could be done by imposing a $\rho-$concavity constraint on the
conditional density of the dependent variable. Using a relatively
low value of $\rho,$ say $-1$ or $-2,$ could be viewed as similar
to a quasi-concavity constraint.\footnote{Note that quasi-concavity is equivalent to uni-modality if and only
if $d=1.$} Convexity of the optimization problem in this case follows from Theorem
7. 

\pagebreak{}

\section*{Appendix A: Log-Concavity Constraints}

A different approach is necessary for $\rho\rightarrow0,$ which corresponds
to $\log-$concavity. In this case $g=\log(f)$ is constrained to
be concave, and $W_{\gamma}(\mu,g^{1/\rho})$ is defined as 

$\underset{(x,y)\in\mathbb{R}^{2m}}{\max}\;x_{-k}^{T}\exp(g_{-k})+x_{k}\left(m-\sum_{i}\exp(g_{-k,i})\right)+y^{T}\mu$

\begin{equation}
-\gamma\underset{i,j}{\sum}\exp\left((x_{i}+y_{j}-M_{ij})/\gamma\right).
\end{equation}

\noindent The index $k$ can be chosen in the same way as described
above to ensure the objective function is convex. 

The gradient in this case is equal to

\begin{equation}
r_{i}:=\frac{\partial W_{\gamma}(\mu,g^{1/\rho})}{\partial g_{-k,i}}=(x_{i}-x_{k})\exp(g_{-k,i}),
\end{equation}

\noindent and the Hessian is 

\begin{equation}
H:=\triangledown^{2}W_{\gamma}(\mu,g^{1/\rho})=ABA^{T}+C
\end{equation}

\noindent where 
\noindent \begin{center}
\textit{$A:=\left[\begin{array}{cc}
D_{\exp(g)} & -\exp(g)\end{array}\right],$}
\par\end{center}

\begin{center}
$B:=\gamma(D_{\exp(g)}-\psi D_{\mathbf{1}\oslash\mu}\psi^{T})^{-1},$
and
\par\end{center}

\begin{center}
$C:=\begin{array}{c}
D_{\exp(g)}\end{array}D_{x_{-k}-x_{k}}.$
\par\end{center}

\section*{Appendix B: An Approximation based on Alternating Bregman Projections}

Algorithm 1 can be derived using the method of alternating Bregman
projections (MABP), which is also the basis for the algorithm proposed
in this section (Bregman, 1967). Bregman explores a class of divergence
measures defined by
\begin{center}
$D_{\varphi}(x\mid y):=\varphi(x)-\varphi(y)-(x-y)^{T}\triangledown\varphi(y),$
\par\end{center}

\noindent where $\varphi:\mathbb{R}^{m}\rightarrow\mathbb{R}$ is
a convex function. In other words, if $\hat{\varphi}_{y}(x)$ is the
first order Taylor series expansion of $\varphi(\cdot)$ at $y\in\mathbb{R}^{m},$
then $D_{\varphi}(x\mid y)=\varphi(x)-\hat{\varphi}_{y}(x).$ Many
distance measures and divergences can be viewed as Bregman divergences.
For example, squared Euclidian distance can be derived using $\varphi(x):=\left\Vert x\right\Vert ^{2},$
and $\varphi(x):=\sum_{i}x_{i}\log(x_{i})$  results in Kullback-Leibler
divergence.

Bregman (1967) described a way to to minimize $\varphi(\cdot)$ subject
to multiple sets of affine constraints using this divergence measure.
As an example, let's consider\footnote{The equality constraints could be replaced with inequalities. However,
the constraints must have a nonempty intersection, be closed, and
be affine. Bauschke and Lewis (2000) prove that a similar algorithm,
which replaces the requirement that the constraints are affine with
a convexity assumption, also converges to the global minimum. } 
\begin{center}
$\underset{x}{\min}\varphi(x)\text{ subject to: }A_{1}x=b_{1},\;A_{2}x=b_{2}.$
\par\end{center}

\noindent For $l\in\{1,2\},$ let the Bregman projection of $y$ onto
the constraint $A_{l}x=b_{l}$ be denoted by
\begin{center}
$P_{l}(y):=\underset{A_{l}x=b_{l}}{\text{arg min}}\;D_{\varphi(\cdot)}(x\mid y).$
\par\end{center}

\noindent MABP begins by initializing $x^{(0)}$ at $\underset{x}{\text{arg min}}\;\varphi(x)$
and the $i^{th}$ iteration takes $x^{(i-1)}$ as input and defines
$x^{(i)}$ as
\begin{center}
$x^{(i)}\leftarrow P_{2}(P_{1}(x^{(i-1)})).$
\par\end{center}

This is a very efficient way to solve an optimization problem when
there is an analytic solution for one or both of the projections.
For example, the two updates found in Algorithm 1 can be viewed as
Bregman (1967) projections of the coupling $\psi$ onto the constraints
$\psi\mathbf{1}_{m}=\mu_{1}$ and $\psi^{T}\mathbf{1}_{m}=\mu_{0}.$
To approximate $f^{\star},$ we need to solve

\begin{equation}
\underset{\psi}{\min}\;\underset{i,j}{\sum}\psi_{ij}M_{ij}+\gamma\psi_{ij}\log\left(\psi_{ij}\right)\;\;\textnormal{subject to:}
\end{equation}

\begin{equation}
\psi^{T}\mathbf{1}_{m}=\mu,
\end{equation}

\begin{equation}
\psi\mathbf{1}_{m}=\alpha_{i}+\beta_{i}\text{a}_{i},\;\text{and}\;\;(\alpha_{i}+\beta_{i}\text{a}_{i})^{\rho}\leq(\alpha_{j}+\beta_{j}\text{a}_{i})^{\rho}\;\;\forall\;i,j\in\{2,...,m-1\}.
\end{equation}

We can only guarantee that MABP converges to the global minimum if
the constraints are affine. The constraint in (37) is not convex,
so theory cannot provide us with a guarantee that MABP converges.
Regardless, MABP is often employed with reasonable success in nonconvex
cases; for examples, see Bauschke, Borwein, and Combettes (2003) and
the references therein. MABP also performs well in our setting, and
since the output of the algorithm will only be used to initialize
Algorithm 2, inaccuracies in the output will not impact our final
density estimates. 

The Bregman divergence corresponding to the objective function in
$(3)$ is
\begin{center}
$D_{W_{\gamma}(\cdot)}(\psi_{ij}\mid\bar{\psi}_{ij})=\gamma\underset{i,j}{\sum}\psi_{i,j}\log\left(\frac{\psi_{ij}}{e\bar{\psi}_{ij}}\right)+\bar{\psi}_{ij}.$
\par\end{center}

\noindent As previously mentioned, the Bregman projection onto the
constraint in (36) is $v\leftarrow\mu\oslash(K^{T}w).$ The constraints
given by (37) can be combined to define the projection,

$P_{2}(\bar{\psi}):=\underset{\psi,\alpha,\beta}{\text{arg min}}\;\underset{i,j}{\sum}\psi_{i,j}\log\left(\frac{\psi_{ij}}{e\bar{\psi}_{ij}}\right)\;\;\text{subject to}\;$

\begin{equation}
\psi\mathbf{1}_{m}=\alpha_{i}+\beta_{i}\text{a}_{i},\;\text{and}\;\;(\alpha_{i}+\beta_{i}\text{a}_{i})^{\rho}\leq(\alpha_{j}+\beta_{j}\text{a}_{i})^{\rho}\;\;\forall\;i,j\in\{2,...,m-1\}.
\end{equation}

Rather than attempting to solve (38) numerically, we can use the change
of variable $f=g^{1/\rho},$ as in Section 3. The following problem
has Kuhn-Tucker conditions that are equivalent to (38) but provide
a reduction in dimensionality. 
\begin{center}
$\underset{g,\alpha,\beta}{\text{arg min}}\;\underset{i}{\sum}g_{i}^{1/\rho}\log\left(\frac{g_{i}^{1/\rho}}{e\bar{v}_{i}}\right)\;\;\text{subject to:}\;$
\par\end{center}

\begin{center}
$g_{i}=\alpha_{i}+\beta_{i}\text{a}_{i}\;\text{and}\;\;\alpha_{i}+\beta_{i}\text{a}_{i}\leq\alpha_{j}+\beta_{j}\text{a}_{i}\;\;\forall\;i,j\in\{2,...,m-1\},$
\par\end{center}

\noindent where $\bar{v}_{i}:=\sum_{j}\bar{\psi}_{ij}.$\footnote{Note that after $v$ is defined using $v\leftarrow\mu\oslash(Kw),$
the equality $\psi=D_{w}KD_{v}$ from the second section implies that
solving for the optimal density is equivalent to solving for the optimal
value of $w$ such that the density $D_{w}Kv$ satisfies the shape
constraint. The variable $\bar{v}$ is $Kv,$ the component of $f$
that is already fixed by the $v-$update. } 

To ensure this optimization problem is convex we need to have $(\rho-1)\log(g_{i}^{1/\rho}/\bar{v}_{i})\leq1$
for every $i\in\{1,2,...,m\}.$ As discussed in Section 2, $v$ and
$w$ are only unique up to a multiplicative constant, so this can
easily be achieved by dividing $v$ by $c\in\mathbb{R}$ and multiplying
$w$ by $c$ in iterations in which this inequality may fail to hold.
The pseudocode for this method is given in Algorithm 3. In this implementation
we renormalize $v$ and $w$ whenever $\max_{i}\;(\rho-1)\log(g_{i}^{1/\rho}/\bar{v}_{i})>3/4,$
and define $c$ as $2^{sgn(\rho)}.$ Generally five to thirty iterations
are sufficient to provide a good initialization for Algorithm 2. 
\begin{algorithm}[H]
\textbf{function }\texttt{MABP}$(\mu,K,\rho)$

$w\leftarrow\mathbf{1}_{m}$

$f\leftarrow\mu$ 

$c\leftarrow2^{sgn(\rho)}$ 

\textbf{for $i=1,2,...$:}

$\;\;\;\;$$v\leftarrow\mu\oslash(Kw)$

$\;\;\;\;$$\bar{v}\leftarrow Kv$

$\;\;\;\;$\textbf{if} $\max_{i}\;(\rho-1)\log(f_{i}/\bar{v}_{i})>3/4$\textbf{:
$v\leftarrow v/c,$ $w\leftarrow cw$ }

$\;\;\;\;g\leftarrow\underset{g}{\text{arg min}}\;\underset{i}{\sum}g_{i}^{1/\rho}\log\left(\frac{g_{i}^{1/\rho}}{e\bar{v}_{i}}\right)\;\;\text{s.t.}\;g\in\mathcal{K}$

$\;\;\;\;f\leftarrow g^{1/\rho}$

$\;\;\;\;$$w\leftarrow f\oslash(Kv)$

$f\leftarrow mf/(f^{T}\mathbf{1})$

\textbf{return $f$}

\caption{Produces an approximate shape-constrained density estimate using MABP.
Note that no constraint regarding the mass of $f$ was made. The mass
of $f$ will be correct when the algorithm converges due to the assignment
$v\leftarrow\mu\oslash(Kw),$ but renormalization at the end of the
algorithm is necessary in the absence of convergence. }
\end{algorithm}

\pagebreak{}

\section*{References}

Afriat, S. N. (1972). Efficiency estimation of production functions.
\textit{International }

\textit{Economic Review}, 568-598. 

\noindent Anderson, T. W., \& Darling, D. A. (1952). Asymptotic theory
of certain ``goodness

of fit'' criteria based on stochastic processes. \textit{The annals
of mathematical }

\textit{statistics},\textit{ }193-212.

\noindent Andrews, D. W. (2000). Inconsistency of the bootstrap when
a parameter is on 

the boundary of the parameter space. \textit{Econometrica, 68}(2),
399-405.

\noindent Bagnoli, M., \& Bergstrom, T. (2005). Log-concave probability
and its applications. 

\textit{Economic Theory, 26}(2), 445-469.

\noindent Bajari, P., Houghton, S., \& Tadelis, S. (2006). Bidding
for incomplete contracts: An 

empirical analysis. \textit{National Bureau of Economic Research.}

\noindent Benamou, J. D., Carlier, G., Cuturi, M., Nenna, L., \& Peyré,
G. (2015). Iterative 

Bregman projections for regularized transportation problems. \textit{SIAM
Journal on }

\textit{Scientific Computing, 37}(2), A1111-A1138.

\noindent Bauschke, H. H., Borwein, J. M., \& Combettes, P. L. (2003).
Bregman monotone 

optimization algorithms. \textit{SIAM Journal on Control and Optimization,
42}(2), 

596-636. 

\noindent Bauschke, H. H., \& Lewis, A. S. (2000). Dykstras algorithm
with bregman 

projections: A convergence proof. \textit{Optimization, 48}(4), 409-427.

\noindent Bregman, L. M. (1967). The relaxation method of finding
the common point of convex 

sets and its application to the solution of problems in convex programming.\textit{
USSR }

\textit{Computational Mathematics and Mathematical Physics, 7}(3),
200-217.

\noindent Carroll, R. J., Delaigle, A., \& Hall, P. (2011). Testing
and estimating shape-constrained 

nonparametric density and regression in the presence of measurement
error. 

\textit{Journal of the American Statistical Association, 106}(493),
191-202.

\noindent Chen, Y., \& Samworth, R. J. (2013). Smoothed $\log-$concave
maximum likelihood 

estimation with applications. \textit{Statistica Sinica}, 1373-1398.

\noindent Cule, M., Samworth, R., \& Stewart, M. (2010). Maximum likelihood
estimation of a 

multi-dimensional log-concave density. \textit{Journal of the Royal
Statistical Society: }

\textit{Series B (Statistical Methodology), 72}(5), 545-607.

\noindent Cuturi, M. (2013). Sinkhorn distances: Lightspeed computation
of optimal transport. 

\textit{Advances in Neural Information Processing Systems}, 2292-2300.

\noindent Cuturi, M., \& Doucet, A. (2014). Fast computation of Wasserstein
barycenters. 

In \textit{International Conference on Machine Learning} (pp. 685-693).

\noindent Cuturi, M., \& Peyré, G. (2016). A smoothed dual approach
for variational 

Wasserstein problems. \textit{SIAM Journal on Imaging Sciences, 9}(1),
320-343.

\noindent Doss, C. R., \& Wellner, J. A. (2016). Global rates of convergence
of the MLEs of log-

concave and s-concave densities. \textit{The Annals of Statistics,
44}(3), 954-981.

\noindent Dümbgen, L., Samworth, R., \& Schuhmacher, D. (2011). Approximation
by log-

concave distributions, with applications to regression. \textit{The
Annals of Statistics, }

\textit{39}(2), 702-730.

\noindent Dümbgen, L., \& Rufibach, K. (2009). Maximum likelihood
estimation of a log-

concave density and its distribution function: Basic properties and
uniform 

consistency. \textit{Bernoulli, 15}(1), 40-68.

\noindent Einmahl, U., \& Mason, D. M. (2005). Uniform in bandwidth
consistency of kernel-

type function estimators. \textit{The Annals of Statistics, 33}(3),
1380-1403.

\noindent Ewerhart, C. (2013). Regular type distributions in mechanism
design and $\rho-$concavity. 

\textit{Economic Theory, 53}(3), 591-603.

\noindent Fan, J. Y., \& Yuan, Y. X. (2001). A new trust region algorithm
with trust region 

radius converging to zero. In \textit{Proceeding of the 5th International
Conference on }

\textit{Optimization: Techiniques and Applications} (pp. 786-794).
ICOTA, Hong Kong.

\noindent Galichon, A. (2016). \textit{Optimal Transport Methods in
Economics}. Princeton University 

Press.

\noindent Grenander, U. (1956). On the theory of mortality measurement.
\textit{Scandinavian }

\textit{Actuarial Journal, 1956}(2), 125-153.

\noindent Guerre, E., Perrigne, I., \& Vuong, Q. (2000). Optimal nonparametric
estimation of 

first-price auctions. \textit{Econometrica, 68}(3), 525-574.

\noindent Hickman, B. R., \& Hubbard, T. P. (2015). Replacing sample
trimming with boundary 

correction in nonparametric estimation of first-price auctions. \textit{Journal
of }

\textit{Applied Econometrics, 30}(5), 739-762.

\noindent Hoffleit, E. D., \& Warren Jr, W. H. (1991). Yale Bright
Star Catalog. \textit{New Haven: }

\textit{Yale Univ. Obs.}

\noindent Horn, R. A., \& Johnson, C. R. (1990). \textit{Matrix analysis}.
Cambridge university press.

\noindent Horowitz, J. L., \& Lee, S. (2017). Nonparametric estimation
and inference under 

shape restrictions. \textit{Journal of Econometrics, 201}(1), 108-126.

\noindent Kantorovitch, L. (1958). On the translocation of masses.
\textit{Management Science, }

\textit{5}(1), 1-4.

\noindent Karlin, S. (1968). \textit{Total positivity}. Stanford:
Stanford University Press.

\noindent Karunamuni, R. J., \& Mehra, K. L. (1991). Optimal convergence
properties of kernel 

density estimators without differentiability conditions. \textit{Annals
of the Institute of }

\textit{Statistical Mathematics, 43}(2), 327-346.

\noindent Kiefer, J., \& Wolfowitz, J. (1956). Consistency of the
maximum likelihood estimator 

in the presence of infinitely many incidental parameters. \textit{The
Annals of }

\textit{Mathematical Statistics}, 887-906.

\noindent Kim, A. K., \& Samworth, R. J. (2016). Global rates of convergence
in $\log-$concave 

density estimation. \textit{The Annals of Statistics, 44}(6), 2756-2779.

\noindent Koenker, R., \& Mizera, I. (2010). Quasi-concave density
estimation. \textit{The Annals of}

\textit{Statistics}, 2998-3027.

\noindent Laffont, J. J., Ossard, H., \& Vuong, Q. (1995). Econometrics
of first-price auctions. 

\textit{Econometrica: Journal of the Econometric Society}, 953-980.

\noindent Krupp, R. S. (1979). Properties of Kruithof's projection
method. \textit{Bell Labs Technical }

\textit{Journal, 58}(2), 517-538.

\noindent Mallows, C. L. (1972). A note on asymptotic joint normality.
\textit{The Annals of }

\textit{Mathematical Statistics}, 508-515.

\noindent Macdonell, W. R. (1902). On criminal anthropometry and the
identification of 

criminals.\textit{ Biometrika, 1}(2), 177-227.

\noindent McAfee, R. P., \& McMillan, J. (1987). Auctions and bidding.
\textit{Journal of economic }

\textit{literature, 25}(2), 699-738.

\noindent Myerson, R. B. (1981). Optimal auction design. \textit{Mathematics
of operations research, }

\textit{6}(1), 58-73.

\noindent Myerson, R. B., \& Satterthwaite, M. A. (1983). Efficient
mechanisms for bilateral 

trading. \textit{Journal of economic theory, 29}(2), 265-281.

\noindent Parzen, E. (1962). On estimation of a probability density
function and mode. \textit{The }

\textit{Annals of Mathematical Statistics, 33}(3), 1065-1076.

\noindent Seregin, A., \& Wellner, J. A. (2010). Nonparametric estimation
of multivariate 

convex-transformed densities. \textit{Annals of statistics, 38}(6),
3751.

\noindent Scott, D. W. (1992). \textit{Multivariate Density Estimation}.\textit{
}Wiley.

\noindent Sheather, S. J., \& Jones, M. C. (1991). A reliable data-based
bandwidth selection 

method for kernel density estimation. \textit{Journal of the Royal
Statistical Society,} 

683-690.

\noindent Silverman, B. W. (1986).\textit{ Density Estimation for
Statistics and Data Analysis}. 

CRC press.

\noindent Sinkhorn, R. (1967). Diagonal equivalence to matrices with
prescribed row and 

column sums. \textit{The American Mathematical Monthly, 74}(4), 402-405.

\noindent Smirnov, N. (1948). Table for estimating the goodness of
fit of empirical distributions. 

\textit{The annals of mathematical statistics, 19}(2), 279-281.

\noindent Solomon, J., De Goes, F., Peyré, G., Cuturi, M., Butscher,
A., Nguyen, A., Du, T.,

\& Guibas, L. (2015). Convolutional Wasserstein distances: Efficient
optimal 

transportation on geometric domains. \textit{ACM Transactions on Graphics
(TOG), }

\textit{34}(4), 66.

\noindent Student. (1908). The probable error of a mean. \textit{Biometrika,
}1-25.

\noindent Van der Vaart, A. W. (2000). \textit{Asymptotic statistics}
(Vol. 3). Cambridge university 

press.

\noindent Villani, C. (2003). \textit{Topics in optimal transportation}
(No. 58). American 

Mathematical Society.

\noindent Wellner, J. A. \& Laha, N. (2017). Bi-s-concave distributions.
\textit{arXiv preprint.}
\end{document}